\documentclass[12pt,a4paper]{article} 
\usepackage{style}
\usepackage{amsmath}
\usepackage{amssymb}
\usepackage{amsthm}
\usepackage{psfrag}
\usepackage{graphicx}
\usepackage{hyperref}
\usepackage{feynmp}
\usepackage{color}
\usepackage{bm}

\renewcommand{\d}{\partial}

\newcommand{\be}{\begin{equation}}
\newcommand{\ee}{\end{equation}}
\newcommand{\beqa}{\begin{eqnarray}}
\newcommand{\eeqa}{\end{eqnarray}}
\newcommand{\bsm}{\begin{smallmatrix}}
\newcommand{\esm}{\end{smallmatrix}}

\renewcommand\k{{\bf k}}

\newcommand{\hMpc}{h\text{Mpc}^{-1}}

\def\d{\partial}
\newcommand{\bseq}{\begin{subequations}}
\newcommand{\eseq}{\end{subequations}}

\renewcommand{\k}{{\bf k}}

\newcommand\vk{\varkappa}

\newcommand{\kmax}{k_{\rm max}}

\renewcommand{\d}{\partial}

\newcommand{\astfootnote}[1]{%
\let\oldthefootnote=\thefootnote%
\setcounter{footnote}{0}%
\renewcommand{\thefootnote}{\fnsymbol{footnote}}%
\footnote{#1}%
\let\thefootnote=\oldthefootnote%
}

\usepackage{color}
\usepackage[caption=false]{subfig}

\renewcommand{\vec}[1]{\boldsymbol{#1}}

\newcommand{\Mpch}{h^{-1}\mathrm{Mpc}}
\renewcommand{\hMpc}{h\,\mathrm{Mpc}^{-1}}
\newcommand{\delD}[1]{(2\pi)^3\delta_\mathrm{D}\left({#1}\right)}
\newcommand{\av}[1]{\left\langle{#1}\right\rangle} 
\renewcommand{\vk}{\vec k}

\newcommand{\vq}{{\vec q}}
\newcommand{\vx}{\vec x}

\newcommand{\hn}{\hat{\vec n}}
\newcommand{\hq}{\hat{\vec q}}
\newcommand{\hk}{\hat{\vec k}}

\usepackage[dvipsnames]{xcolor}
\newcommand{\bb}[1]{\textcolor{purple}{#1}}
\def\cG{\mathcal{G}}

\usepackage{color}
\def\beq{\begin{eqnarray}}
\def\eeq{\end{eqnarray}}

\relax

\title{Cosmology with the Redshift-Space Galaxy Bispectrum Monopole at One-Loop Order}

\author[a,b]{Oliver H.\,E. Philcox,\footnote{\href{mailto:ohep2@cantab.ac.uk}{ohep2@cantab.ac.uk}}}
\author[b]{Mikhail M. Ivanov,\footnote{\href{mailto:ivanov@ias.edu}{ivanov@ias.edu}, Einstein Fellow}}
\author[b]{Giovanni Cabass,}
\author[c]{\\Marko Simonovi\'c,}
\author[b]{Matias Zaldarriaga,}
\author[d,e]{and Takahiro Nishimichi}

\affiliation[a]{Department of Astrophysical Sciences, Princeton University,\\ Princeton, NJ 08540, USA}%
\affiliation[b]{School of Natural Sciences, Institute for Advanced Study,\\1 Einstein Drive, Princeton, NJ 08540, USA}
\affiliation[c]{Theoretical Physics Department, CERN,\\1 Esplanade des Particules, Geneva 23, CH-1211, Switzerland}
\affiliation[d]{Center for Gravitational Physics and Quantum Information, \\ Yukawa Institute for Theoretical Physics, Kyoto University, Kyoto 606-8502, Japan}
\affiliation[e]{Kavli Institute for the Physics and Mathematics of the Universe (WPI), UTIAS \\The University of Tokyo, Kashiwa, Chiba 277-8583, Japan}

\abstract{\small
We study the cosmological information content of the redshift-space galaxy bispectrum monopole at one-loop order in perturbation theory. 
We incorporate all effects necessary for comparison to data: fourth-order galaxy bias, infrared resummation (accounting for the non-linear evolution of baryon acoustic oscillations), ultraviolet counterterms, non-linear redshift-space distortions, stochastic contributions, projection, and binning effects. The model is implemented using FFTLog, and validated with the PT Challenge suite of $N$-body simulations, whose large volume
allows for high-precision tests. Focusing on the mass fluctuation amplitude, $\sigma_8$, and galaxy bias parameters, we find that including one-loop corrections allow us to significantly extend the range of scales over which the bispectrum can be modeled, and greatly tightens constraints on bias parameters. However, this does not lead to noticeable improvements in the $\sigma_8$ errorbar due to the necessary marginalization over a large number of nuisance parameters with conservative
priors. Analyzing a BOSS-volume likelihood, we find that the addition of the one-loop bispectrum may lead to improvements on primordial non-Gaussianity constraints by $\lesssim 30\%$ and on $\sigma_8$ by $\approx 10\%$, though we caution that this requires pushing the analysis to short-scales where the galaxy bias parameters may not be correctly recovered; this may lead to biases in the recovered parameter values. We conclude that restrictive priors from simulations or higher-order statistics such as the bispectrum multipoles will be needed in order to realize the full information content of the galaxy bispectrum.
}

\normalsize

\begin{document}

\begin{flushright}
YITP-22-60, CERN-TH-2022-092
\end{flushright}

\maketitle

\section{Introduction}

The distribution of matter in the Universe is not Gaussian. As such, any optimal analysis of large-scale structure surveys must involve non-Gaussian statistics. Historically, the distribution of galaxies has been analyzed by way of the power spectrum, which is a complete statistic only in the Gaussian (large-scale) limit. Such analyses have seen particular successes, both in measuring the Universe's growth rate and expansion history via baryon acoustic oscillation signatures \citep[e.g.,][]{Grieb:2016uuo,Alam:2016hwk,Satpathy:2016tct,Beutler:2016ixs,Beutler:2016arn,deMattia:2020fkb,BOSS:2016wmc,eBOSS:2020yzd}, and in the direct determination of cosmological parameters using full-shape methods \citep[e.g.,][]{Percival:2006gt,Ivanov:2019pdj,Ivanov:2019hqk,Ivanov:2021zmi,Ivanov:2020ril,Philcox:2020vvt,Philcox:2021kcw,Chudaykin:2020ghx,Philcox:2020xbv,Philcox:2022sgj,DAmico:2019fhj,DAmico:2020tty,DAmico:2020ods,DAmico:2020kxu,Colas:2019ret,Wadekar:2020hax,Kobayashi:2021oud,Chen:2021wdi,Chen:2022jzq}. However, if we wish to extract more information from the observational data-sets (and thus enhance their utility at fixed experimental cost), we must additionally include higher-order statistics, of which the simplest are the bispectrum and three-point correlation function.

Much as the power spectrum describes the distribution of pairs of points, the bispectrum encodes the statistics of triplets \citep{1980lssu.book.....P}. Despite being a subject of significant early work \citep[e.g.,][]{1975ApJ...196....1P,1977ApJ...217..385G,Feldman:2000vk,Marin:2013bbb,Scoccimarro:2000sp,Scoccimarro:1997st,Sefusatti:2006pa,Scoccimarro:2000sn,Scoccimarro:1999ed}, the bispectrum has been rarely used in modern cosmological analyses (though see \citep{Gil-Marin:2014baa,Gil-Marin:2014sta,Gil-Marin:2016wya}), owing to difficulties in its measurement \citep[c.f.,][]{Scoccimarro:2015bla,Philcox:2019hdi,Philcox:2020xod}, modeling, and interpretation. On the other hand, its potential utility has been demonstrated a number of times, with claims of tight constraints on a number of cosmological parameters, particularly those encoding extensions to the $\Lambda$CDM paradigm, such as neutrino masses, primordial non-Gaussianity 
and the breaking of the equivalence principle
\citep{Baldauf:2016sjb,Hahn:2019zob,Welling:2016dng,MoradinezhadDizgah:2018ssw,Hahn:2020lou,MoradinezhadDizgah:2020whw,Ruggeri:2017dda,Song:2015gca,Karagiannis:2018jdt,Peloso:2013zw,Kehagias:2013yd,Valageas:2013cma,Creminelli:2013mca,Creminelli:2013poa,Creminelli:2013nua,Lewandowski:2019txi,Yankelevich:2018uaz,Crisostomi:2019vhj,Oddo:2019run,Oddo:2021iwq,Chudaykin:2019ock,Rizzo:2022lmh}. To fully utilize this information, we require theoretical models capable of predicting the shape of the bispectrum and its dependence on the parameters of interest. This has been a subject of significant work, starting from the matter bispectrum \citep{Baldauf:2014qfa,Angulo:2014tfa,Lazanu:2018yae,Alkhanishvili:2021pvy,2010MNRAS.406.1014S,Valageas_2011,Takahashi_2020,deBelsunce:2018xtd}, then incoroporating the effects of redshift-space distortions \citep{DiDio:2018unb,Gil-Marin:2014pva,Scoccimarro:1999ed,Desjacques:2018pfv} and galaxy bias \citep{Smith:2007sb,Nadler:2017qto,Eggemeier:2018qae,Ivanov:2021kcd,Oddo:2019run,Eggemeier:2021cam}, most successfully using the Effective Field Theory of Large Scale Structure (hereafter EFTofLSS, \citep{Carrasco:2012cv,Baumann:2010tm}, see \citep{Cabass:2022avo} for a recent review).

Whilst the above references have been pivotal to the development of a bispectrum model, few contain all the necessary ingredients to allow for robust comparison of theory and observation. In particular, one must account for the backreaction of short-scale physics on the large-scale bispectrum \citep{Baldauf:2014qfa,Angulo:2014tfa,Steele:2020tak}, long-wavelength displacements \citep{Ivanov:2018gjr,Blas:2016sfa,Senatore:2017pbn,Lewandowski:2018ywf,Senatore:2014via,Vasudevan:2019ewf,Baldauf:2015xfa,Vlah:2015zda}, and survey geometry \citep{Philcox:2021ukg,Pardede:2022udo}, all of which can lead to biases in derived parameters if not properly accounted for. In \citep{Ivanov:2021kcd} a complete model for the tree-level (leading-order) bispectrum of galaxies in redshift-space was presented and validated, including all the above effects (see also \citep{DAmico:2019fhj}).  This allows for precise modelling of the angle-averaged bispectrum monopole, and has facilitated a number of analyses constraining $\Lambda$CDM parameters  \citep{Philcox:2021kcw} and primordial non-Gaussianity \citep{Cabass:2022wjy,Cabass:2022ymb}. However, this model was restricted to relatively large scales ($k < 0.08~\Mpch$ at $z = 0.61$). If we wish to further exploit the constraining power of the bispectrum, we must push to smaller scales, by extending the perturbation theory to next order. Whilst \citep{DAmico:2022gki} has recently demonstrated some work in this direction, a full model for the one-loop bispectrum (including all relevant phenomena such as projection effects) has not yet been presented and validated with simulations.

In this work, we present a complete and systematic computation of the redshift-space galaxy bispectrum monopole at one-loop order.
This includes all effects necessary to compare with observational data: deterministic contributions, counterterms, bias renormalization, stochasticity, bin-averaging, and coordinate distortions. This involves the galaxy density at fourth-order: we systematically account for all bias operators (following \citep{Eggemeier:2018qae}), and include full treatment of all necessary redshift-space counterterms, ensuring a convergent Taylor series. Our model necessarily depends on a number of free parameters: these account for the unknown complexities of ultraviolet physics (such as galaxy formation physics and feedback), and ensure physical robustness. Efficient computation of the one-loop bispectrum is non-trivial; as such, we devote a significant portion of this work to discussing its practical computation with the FFTLog algorithm \citep{Simonovic:2017mhp}. We compare the theoretical predictions to real- and redshift-space bispectra obtained from the PT Challenge simulations \citep{Nishimichi:2020tvu}, which serve both to validate the approach and to assess the information content of the one-loop bispectrum model. Though we restrict to the measurement of $\sigma_8$ and primordial non-Gaussianity parameters, one can constrain a variety of other phenomena with the bispectrum, and, further still, our methodology can be extended to other correlators including the bispectrum multipoles \citep{Rizzo:2022lmh,Gualdi:2020ymf} and the recently-detected trispectrum \citep{Philcox:2021hbm,Gualdi:2022kwz}.

\vskip 8 pt

The remainder of this paper is structured as follows. The theoretical model is presented in \S\ref{sec: theory}, before its implementation is outlined in \S\ref{sec: implementation}. In \S\ref{sec: data} we give details of the data and analysis choices used to validate the model, before presenting the results of likelihood analyses using the real- and redshift-space galaxy bispectrum in \S\ref{sec: results-real}\,\&\,\ref{sec: results-rsd} respectively. \S\ref{sec: boss} comments on the method's applicability to current datasets, with a summary and discussion given in \S\ref{sec: discussion}. Finally, various technical details are presented in the Appendices, including: \ref{app:kernels} the perturbation theory kernels, \ref{app:fftlog} details of the bispectrum integration routines, \ref{app:ctrs} discussion of the redshift-space counterterms, and \ref{app:stoch} derivation of the stochastic bispectrum components. Appendix~\ref{sec:proj}
is devoted to prior volume effects.
 The key plots of this work are Fig.\,\ref{fig: bk-plot}, showing the one-loop bispectrum components, and Fig.\,\ref{fig:boss-like}, displaying the utility of the bispectrum for a BOSS-like survey.

\section{Theoretical Model for the One-Loop Bispectrum}\label{sec: theory}

In this work, we analyze the power spectrum and bispectrum of biased tracers (\textit{i.e.}\ galaxies) in redshift space at one-loop order. Whilst the one-loop power spectrum and tree-level bispectrum have been described in detail before \citep[e.g.][]{Ivanov:2021kcd,DAmico:2019fhj}, a complete model for the one-loop bispectrum has not been presented before (though some aspects can be found in \citep{DAmico:2022gki}) and will be discussed below, with additional technical details found in the appendices.
Here, we will restrict to Gaussian initial conditions; extension to primordial non-Gaussianity is discussed in \S\ref{subsec: png}.

In the EFTofLSS, the bispectrum is comprised of the following terms at one-loop order \citep[e.g.,][]{Bernardeau:2001qr,Baldauf:2014qfa,Steele:2020tak}:
\beq
    B_{\rm 1-loop}(\vk_1,\vk_2,\vk_3) = B_{211} + \left[B_{222}+B^I_{321} + B^{II}_{321}+B_{411}\right] + B_{\rm ct} + B_{\rm stoch},
\eeq
where the first and second terms give the tree-level and one-loop bispectrum in Eulerian perturbation theory, $B_{\rm ct}$ is the derivative and counterterm contribution, and $B_{\rm stoch}$ encodes stochasticity. This is strictly a function of five variables: three lengths, $\{k_1, k_2, k_3\}$ and two angles, $\{\mu_1, \mu_2\}$, for $\mu_i\equiv \hk_i\cdot\hn$ with line-of-sight $\hn$ (hereafter LoS), noting that $k_1\mu_1+k_2\mu_2+k_3\mu_3=0$. In real-space, this reduces to just three variables: $\{k_1,k_2,k_3\}$.

\subsection{Bias Expansion}
To compute the bispectrum within Eulerian perturbation theory, our first step is to express real-space galaxy density field, $\delta_g$, in terms of a basis of \textit{bias operators}, \textit{i.e.}\ all combinations of the density and velocity fields ($\delta$ and $\theta$) consistent with the relevant symmetries up to a given order in perturbation theory \cite{McDonald:2009dh,Assassi:2014fva,Angulo:2015eqa,Senatore:2014eva,Mirbabayi:2014zca,Desjacques:2016bnm,Eggemeier:2018qae}. 
For the one-loop bispectrum, we require terms up to fourth-order ($\delta_L^4$), and here use the basis of Galileon operators proposed in \citep{Eggemeier:2018qae}:
\beq\label{eq: bias-expan-4th}
    \delta_g &=& \left\{\bb{b_1}\delta\right\} + \left\{\frac{\bb{b_2}}{2}\delta^2+\bb{\gamma_2}\,\cG_2(\Phi_v)\right\}\\\nonumber
    &&\,+\,\left\{\frac{\bb{b_3}}{6}\delta^3+\bb{\gamma_2^\times}\,\delta\,\cG_2(\Phi_v)+\bb{\gamma_3}\,\cG_3(\Phi_v)+\bb{\gamma_{21}}\,\cG_2(\varphi_2,\varphi_1)\right\}\\\nonumber
    &&\,+\,\left\{\bb{\gamma_{21}^\times}\,\delta\,\cG_2(\varphi_2,\varphi_1)+\bb{\gamma_{211}}\,\cG_3(\varphi_2,\varphi_1,\varphi_1)+\bb{\gamma_{22}}\,\cG_2(\varphi_2,\varphi_2)+\bb{\gamma_{31}}\,\cG_2(\varphi_3,\varphi_1)\right\}+\mathcal{O}(\delta^5) ,
\eeq
where curly brackets separate operators of different order and the bias parameters are marked in color. In \eqref{eq: bias-expan-4th}, we drop any terms that do not appear in the one-loop bispectrum; these are all composite local evolution operators such as $\delta^4$ and $\delta^2\,\cG_2(\Phi_v)$. Here we have ignored both higher-derivative operators (which we return to below) and bias renormalization, which is discussed in Appendix \ref{app:fftlog}. The Galileon operators are defined by
\beq
    \cG_2(\Phi_v) &\equiv& \nabla_i\nabla_j \Phi_v\nabla^i\nabla^j\Phi_v-(\nabla^2\Phi_v)^2 , \\\nonumber
    \cG_3(\Phi_v) &\equiv& 2\nabla_i\nabla_j\Phi_v\nabla^j\nabla_k\Phi_v\nabla^k\nabla^i\Phi_v-3\nabla_i\nabla_j\Phi_v\nabla^i\nabla^j\Phi_v\nabla^2\Phi_v+(\nabla^2\Phi_v)^3,
\eeq
where $\Phi_v\equiv \nabla^{-2}\theta$ is the velocity potential, equal to the Newtonian potential $\Phi \equiv \nabla^{-2}\delta$ at leading order. These can be simply generalized to functions of multiple potentials, with \eqref{eq: bias-expan-4th} involving the LPT potentials $\varphi_{1,2}$, satisfying
\beq
    \nabla^2\varphi_1 = -\delta, \qquad \nabla^2\varphi_2 = -\cG_2(\varphi_1).
\eeq
Up to third order, this is equivalent to the bias expansion used in \citep{Assassi:2014fva} and previous works \citep[e.g.,][]{Ivanov:2021kcd,Ivanov:2019pdj}, with the relations\footnote{This is obtained by noting that $\cG_2(\Phi_v)$ is just the usual tidal tensor $\cG_2\equiv-\cG_2(\delta)$ at first order, and that $\cG_2(\varphi_2,\varphi_1)$ is a combination of the $\cG_2$ and $\Gamma_3$ operators of \citep{Assassi:2014fva}.}
\beq\label{eq: bias-relation}
\bb{\gamma_2}\equiv b_{\cG_2},\quad  \bb{\gamma_{21}}\equiv -\tfrac{4}{7}\left(b_{\cG_2}+b_{\Gamma_3}\right).
\eeq

Utilizing \eqref{eq: bias-expan-4th}, and expanding each operator in terms of the linear density field $\delta^{(1)}\equiv\delta_L$, we can define the $n$-th order contributions to the galaxy density field:
\beq\label{eq: delta-n-exp}
    \delta^{(n)}_g(\vk) = \int_{\vq_1\ldots\vq_n}\delD{\vq_1\cdots+\vq_n-\vk}K_n(\vq_1,\cdots,\vq_n)\delta_L(\vq_1)\cdots\delta_L(\vq_n),
\eeq
where the real-space kernels $K_n$ are given in Appendix \ref{app:kernels-real} and depend on the bias parameters given above. Furthermore, this generalizes to the redshift-space density field, $\delta_s(\vk)$, using the well-known mapping \citep[e.g.,][]{Kaiser:1987qv}
\beq\label{eq: rsd-mapping}
    \delta_g^{(s)}(\vk) = \delta_g(\vk) + \int d\vx\,e^{-i\vk\cdot\vx}\left[e^{-ik_z f u_z(\vx)}-1\right]\left(1+\delta_g(\vx)\right),
\eeq
where $f$ is the logarithmic growth rate, $u_z(\vq) = (i\mu_{\vq}/q)\theta(\vq)$ is the Fourier-space LoS velocity field, and $\mu_{\vq}\equiv\hat{\vq}\cdot\hn$, for LoS vector $\hn$. The associated kernels, analogous to \eqref{eq: delta-n-exp}, are labelled as $Z_n$ and defined for $n\leq 4$ in Appendix \ref{app:kernels-rsd}. 

\subsection{Deterministic Contributions}
Utilizing the redshift-space kernels of Appendix \ref{app:kernels-rsd}, the tree-level bispectrum, $B_{211}\equiv\av{\delta^{(2)}_g\delta^{(1)}_g\delta^{(1)}_g}$ can be written
\beq
    B_{211}(\vk_1,\vk_2,\vk_3) = 2\,Z_2(\vk_1,\vk_2)Z_1(\vk_1)Z_1(\vk_2)P_L(k_1)P_L(k_2) + \text{2 cyc.},
\eeq
where $P_L(k)$ is the linear power spectrum (though see the below discussion on infrared resummation). This depends on the bias parameters $\{b_1,b_2,\gamma_2\}$, as well as the growth rate, $f(z)$. The one-loop terms can be written as loop integrals over the linear power spectrum, and come in four flavors \citep[e.g.,][]{Bernardeau:2001qr,Baldauf:2014qfa}:
\beq\label{eq: Bk-one-loop}
    B_{222}(\vk_1,\vk_2,\vk_3) &=& 8\int_{\vq}Z_2(\vk_1+\vq,-\vq)Z_2(\vk_1+\vq,\vk_2-\vq)Z_2(\vk_2-\vq,\vq)\\\nonumber
    &&\qquad\times\,P_L(q)P_L(|\vk_1+\vq|)P_L(|\vk_2-\vq|), \\\nonumber
    B_{321}^I(\vk_1,\vk_2,\vk_3) &=&  6\,Z_1(\vk_1)P_L(k_1)\int_{\vq}Z_3(-\vq,\vq-\vk_2,-\vk_1)Z_2(\vq, \vk_2-\vq)\\\nonumber
    &&\qquad\times\,P_L(q)P_L(|\vk_2- \vq|)+\text{5 perm.}, \\\nonumber
    B_{321}^{II}(\vk_1,\vk_2,\vk_3) &=& 6\,Z_2(\vk_1,\vk_2)Z_1(\vk_2)P_L(k_1)P_L(k_2)\int_{\vq}Z_3(\vk_1,\vq,-\vq)P_L(q)+\text{5 perm.}, \\\nonumber
    B_{411}(\vk_1,\vk_2,\vk_3) &=& 12\,Z_1(\vk_1)Z_1(\vk_2)P_L(k_1)P_L(k_2)\int_{\vq}Z_4(\vk_1,\vk_2,\vq,-\vq)P_L(q)+\text{2 cyc.},
\eeq
where the $B_{321}^{II}$ spectrum is similar to the $P_{13}(\vk_1)$ contribution to the one-loop power spectrum. Computation of the loop integrals can be performed via explicit numerical integration or with the FFTLog method \citep{Simonovic:2017mhp}; we discuss the latter in \S\ref{sec: implementation}, with details presented in Appendix \ref{app:fftlog}. As well as the tree-level biases, these spectra involve the higher-order parameters $\{b_3,\gamma_2^\times,\gamma_3,\gamma_{21},\gamma_{21}^\times,\gamma_{211},\gamma_{22},\gamma_{31}\}$, of which only $\gamma_{21}$ appears in the one-loop power spectrum.

\subsection{Counterterms}
To ensure a self-consistent theoretical model, we require a set of counterterms, which account for non-idealities in fluid equations (via the viscous stress tensor), and absorb the unknown ultraviolet (UV, $q\gg k$) behavior of the loop integrals in \eqref{eq: Bk-one-loop} \citep[e.g.,][]{Baldauf:2014qfa,Angulo:2014tfa,Steele:2020tak}. 
For the one-loop bispectrum in real-space, these operators are degenerate with derivative operators in the bias expansion, such as $\nabla^2\delta$. Furthermore, the redshift-space bispectrum contains additional counterterms that appear after the renormalization of contract operators in the perturbative mapping of \eqref{eq: rsd-mapping}; these are discussed in detail in Appendix~\ref{app:ctrs}.

The overall bispectrum counterterm contribution 
can be written as
\be\label{eq: Bk-ctr}
 \begin{split}
B_{\rm ctr}(\k_1,\k_2,\k_3)=2Z_1(\k_1)Z_1(\k_2)(F_2^{\rm ctr} + Z_2^{\rm ctr})(\k_1,\k_2)P_{L}(k_1)P_{L}(k_2)
+\text{cyclic perms.} \,,
 \end{split}
\ee
where $F_2^{\rm ctr}(\k_1,\k_2)$ is the real-space counterterm kernel~\citep{Eggemeier:2018qae}:
\beq
    F^{\rm ctr}_2(\vk_1,\vk_2) &=& -\left\{\left[\bb{\beta_{B,a}}(k_1^2+k_2^2)+\bb{\beta_{B,b}}k_3^2\right]F_2(\vk_1,\vk_2)\right.\\\nonumber
    &&\qquad\,+\left.\left[\bb{\beta_{B,c}}(k_1^2+k_2^2)+\bb{\beta_{B,d}}k_3^2\right]\kappa(\vk_1,\vk_2)+\bb{\beta_{B,e}}\,\vk_1\cdot\vk_2\right\}\frac{1}{k_{\rm NL}^2}\,,
\eeq
where $\kappa(\vk_1,\vk_2)=(\hat \vk_1\cdot \hat \vk_2)-1$ (cf.\,\ref{eq: ang-def}), and we choose the non-linear scale $k_{\rm NL}=0.45\hMpc$~\cite{Baldauf:2016sjb,Chudaykin:2020aoj,Chudaykin:2020hbf}. \eqref{eq: Bk-ctr} additionally involves the $\mu$-dependent redshift-space kernel $Z_2^{\rm ctr}$ defined by
\beq\label{eq:rsdcount}
Z_2^{\rm ctr}(\k_1,\k_2) &=& \frac{1}{k_{\rm NL}^2}\Bigg(\bb{C_1} k_{3z}^2 F_2(\k_1,\k_2)  +
\bb{C_2} k_{3z}^2  \left(\mu^2_3-\frac{1}{3}\right) F_2(\k_1,\k_2)   
+ \bb{C_3} k_{3z}^2  \\\nonumber
&&\,+\,\bb{C_4} k_{3z}^2 \left(
\frac{ (\k_1\cdot \k_2) \mu_1 \mu_2 }{k_1  k_2} 
-\frac{\mu_1^2+\mu_2^2}{3}  
+\frac{1}{9}\right) 
+ \bb{C_5} f  k_{3z}^2 \mu_3^2  G_2(\k_1,\k_2) \\\nonumber
&&\,+\,\bb{C_6} f k_{3z}^3 
\left[\frac{\mu_1}{k_1} 
\left(\mu_2
-\frac{1}{3}\right) + \frac{\mu_2}{k_2} 
\left(\mu_1
-\frac{1}{3}\right)\right]
+\bb{C_7} f  k_{3z}^3  \left[\frac{\mu_1}{k_1}+\frac{\mu_2}{k_2} \right]  \\\nonumber
&&\,+\,\bb{C_8} f^2 k_{3z}^4 \frac{\mu_1}{k_1} \frac{\mu_2}{k_2}
 + \bb{C_9} k_{3z}^2 \left(\frac{(\k_1 \cdot \k_2)^2}{k^2_1 k^2_2 }-\frac{1}{3}\right) \Bigg),
\eeq
as derived in Appendix \ref{app:ctrs}, with $k_{3z}\equiv k_3\mu_3$.
In principle, two combinations of $C_1,C_2$ and $C_5$ 
are constrained by the power spectrum, so only one 
parameter out of three is independent here.
In practice, however, we 
did not find any difference between 
imposing the power spectrum
constraints on $C_1,C_2,C_5$ or 
treating them as free parameters.
This is why we proceed 
with keeping them free
in what follows. 
In total, the one-loop bispectrum counterterm depends on 14 free parameters, $\{\beta_{B,i}\}$,
and $\{C_i\}$ in addition to the one-loop power spectrum counterterms.\footnote{Some of the bispectrum counterterms are time
integrals of the power spectrum counterterms, 
but since we do not know the time-dependence, 
we prefer 
to keep all the counterterms free. 
}

Notably, many of the counterterms appearing in \eqref{eq:rsdcount}
are degenerate at the bispectrum monopole level; nevertheless, we prefer to keep all of them in the model, 
and marginalize over them within physically motivated priors.
This is done for two main reasons. First, terms with different powers of $\mu$ can, in principle, be distinguished even at the bispectrum 
monopole level thanks to the Alcock-Paczynski projection effect~\cite{Alcock:1979mp}, which is described below. Second, the degeneracy between these terms can be broken with higher order angular multipole moments of the bispectrum~\cite{Scoccimarro:1999ed,Scoccimarro:2015bla}, which we will analyze in the future.

\subsection{Stochasticity}
Contributions to the bispectrum are also sourced by the non-deterministic part of the density field $\epsilon$~\cite{Assassi:2014fva,Angulo:2015eqa,Senatore:2014eva,Mirbabayi:2014zca,Desjacques:2016bnm}, \textit{i.e.}\ that uncorrelated with $\delta_L$. At tree-level, this gives two terms, $\propto 1/\bar n, P(k)/\bar n$ (arising from Poissonian shot-noise with sample density $\bar n$), whilst at one-loop order, we must keep contributions suppressed by $(k/k_{\rm halo})^2$, where $k^{-1}_{\rm halo}$ is some characteristic halo size. From \citep{Eggemeier:2018qae}, we have the following form at next-to-leading order in real-space:
\beq\label{eq: Bstoch}
    \left.B_{\rm stoch}(\vk_1,\vk_2,\vk_3)\right|_{\rm real-space} &=& \frac{\bb{A_{\rm shot,0}}}{\bar n^2}+
   \frac{\bb{A_{\rm shot,1}}}{\bar n^2}
    \left(k_1^2+k_2^2+k_3^2\right)\\\nonumber
    &&\,+\,\left[\left(\bb{B_{\rm shot}}+
    \bb{S_{1}}\frac{k_1^2}{k_{\rm NL}^2}+\bb{S_0}\frac{k_2^2 + k_3^2}{k_{\rm NL}^2}\right)\frac{P_{L}(k_1)}{\bar n} + \text{2 cyc.}\right],
\eeq
depending on another five free parameters $\{A_{\rm shot,0}, B_{\rm shot}\}$ and
$\{A_{\rm shot,1},S_0,S_1\}$, which cannot be constrained with the one-loop power spectrum. In the Poisson limit, $A_{\rm shot,0}=B_{\rm shot}=1$, with all higher-order terms (arising, for example, from halo exclusion) vanish. 

In redshift-space, significantly more dependencies arise. A systematic derivation of these is presented in Appendix~\ref{app:stoch} and yields the following 
expression: 
\be
B_{\rm stoch}= B^{(P\bar n^{-1})}_{\rm stoch} +B^{(\bar n^{-2})}_{\rm stoch} + B^{(k^2 P\bar n^{-1})}_{\rm stoch} + B^{(k^2 \bar n^{-2})}_{\rm stoch}\,,
\ee
where 
\be
\begin{split}
& B^{(P\bar n^{-1})}_{\rm stoch}(\k_1,\k_2,\k_3)=
\Bigg[
\bb{b_1} \bb{B_{\rm shot}}+ f\mu^2 (1+\bb{P_{\rm shot}})
\Bigg]\frac{Z_1(\k_1)P_{L}(k_1)}{\bar n}+\text{cyc.}
 \\
 & B^{(\bar n^{-2})}_{\rm stoch}(\k_1,\k_2,\k_3)=\frac{\bb{A_{\rm shot,0}}}{\bar n^2}\,,
 \\
& B^{(k^2 P\bar n^{-1})}_{\rm stoch}(\k_1,\k_2,\k_3)=\Bigg[ 
\sum_{n=1,2,3,4} 
\bb{S_{n}} k_1^2 \mu^{2(n-1)}_1 
+ \bb{S_5} Z_1(\k_1) (\mu_2^2 k_2^2+\mu_3^2 k_3^2) \\
 &
+ \bb{S_6} Z_1(\k_1) \mu_1^2(\mu_2^2 k_2^2+\mu_3^2 k_3^2)
 +\bb{S_7}  Z_1(\k_1)\frac{k_{1z}}{k_1^2}(k_{2z}^3+k_{3z}^3) 
 + \bb{S_0} Z_1(\k_1)(k_2^2+k_3^2) \Bigg]\frac{P_{L}(k_1)}{k_{\rm NL}^2\bar n}+\text{cyc.}\,,\\
& B^{(k^2 \bar n^{-2})}_{\rm stoch}(\k_1,\k_2,\k_3)=\left[\bb{A_{\text{shot},1}} (k_1^2+k_2^2+k_3^2)+\bb{A_{\text{shot},2}} (k_{3z}^2+k_{1z}^2+k_{2z}^2)\right]
\frac{1}{k_{\rm NL}^2\bar n^2}\,.
\end{split} 
\ee
This expression shares the parameter $P_{\rm shot}$ with the power spectrum, but includes an additional 12 nuisance coefficients: $\{ \{S_{n}\},\{A_{\rm shot,n}\}, B_{\rm shot}\}$. $P_{\rm shot}$ is defined as a constant rescaling 
of the stochastic power spectrum~\cite{Ivanov:2021kcd}, 
\be
\langle \epsilon(\k)\epsilon(\k')\rangle =(2\pi)^3 \delta_D^{(3)}(\k+\k')\frac{1+P_{\rm shot}}{\bar n}\,.
\ee
Note that in the absence of projection effects the counterterms
$A_{\text{shot},1}$ and $A_{\text{shot},2}$
are fully degenerate. Therefore, for the purposes of this study 
we will set $A_{\text{shot},2}=0$.

\subsection{Infrared Resummation}
An additional complication arises from the effects of long-wavelength displacements, which can be consistently treated using ``infrared resummation''. A rigorous derivation of this was presented in \citep{Blas:2016sfa,Ivanov:2018gjr} in the context of time-sliced perturbation theory~\citep{Blas:2015qsi}, and, at tree-level order, can be implemented by replacing the linear power spectrum $P_L$ with its IR-resummed equivalent, \textit{i.e.}\footnote{Note that the general expressions for the IR resummed functions in terms of differential 
operators presented in 
~\citep{Blas:2016sfa,Ivanov:2018gjr} can be applied to the whole power spectrum,
\textit{i.e.}\ in principle the procedure of these works does not require a wiggly-smooth 
decomposition. We prefer to use this decomposition because it results in a significant time gain.
}
\beq
    P_L(k)&\to& P_{nw}(k)+e^{-\Sigma^2k^2(1+f\mu^2(2+f))-\delta\Sigma^2k^2f^2\mu^2(\mu^2-1)}P_w(k)\\\nonumber
    &\equiv& P_{nw}(k)+e^{-\Sigma^2_{\rm tot}(\mu)k^2}P_w(k),
\eeq
where $P_w$ and $P_{nw}$ are the wiggly and smooth parts of the power spectrum respectively. This has the effect of damping the oscillatory component by a $k$- and $\mu$-dependent factor. The damping scales are given in terms of the broadband power spectrum as
\beq
    \Sigma^2 &=& \frac{1}{6\pi^2}\int_0^{k_S}dq\,P_{nw}(q)\left[1-j_0(qr_{\rm BAO})+2j_2(qr_{\rm BAO})\right], \\\nonumber
    \delta\Sigma^2 &=& \frac{1}{2\pi^2}\int_0^{k_S}dq\,P_{nw}(q)j_2(qr_{\rm BAO}),
\eeq
where $r_{\rm BAO}$ is the sound-horizon scale and $k_S\sim 0.1\hMpc$. At one-loop order, the IR-resummed bispectrum can be written schematically as
\beq
    B_{\rm 1-loop} &=& B_{\rm 211}[P_{nw}+(1+k^2\Sigma^2_{\rm tot})e^{-\Sigma^2_{\rm tot}k^2}P_w]\\\nonumber
    &&\,+\,\left(B_{321}^I+B_{321}^{II}+B_{411}\right)[P_{nw}+e^{-\Sigma^2_{\rm tot}k^2}P_{w}] ,
\eeq
where $B[P]$ indicates that the bispectrum should be evaluated using the power spectrum $P$ and we have dropped the counterterms and stochasticity~\citep{Ivanov:2018gjr}. In this case, the loop corrections become more complex, since the damping factor, $\Sigma^2_{\rm tot}$ is a function of the redshift-space angles $\mu$. To allow for efficient computation via the FFTLog procedure (Appendix \ref{app:fftlog}), we here adopt the isotropic approximation for the one-loop terms, dropping any $\mu$-dependence in $\Sigma^2_{\rm tot}$ inside the integral. This is expected to be a good approximation in practice, and is exact for the real-space case. 
Note that we keep the full redshift-dependent damping function
in the tree-level expressions, \textit{i.e.}\ the isotropic templates are used 
only for the computations of the one-loop corrections.

\subsection{Coordinate Rescalings}
To compare to observations, we must distill the full bispectrum $B(\vk_1,\vk_2,\vk_3)$ to a lower-dimensional form. In this work, we consider the angle-averaged bispectrum monopole (as in \citep{Ivanov:2021kcd,Philcox:2021kcw,Cabass:2022wjy}), defined via
\beq\label{eq: Bk-mon}
    B_0(k_1,k_2,k_3) = \int_{-1}^{1}\frac{d\mu}{2}\int_0^{2\pi}\frac{d\phi}{2\pi}\,B(k_1,k_2,k_3,\mu_1[\mu],\mu_2[\mu,\phi]),
\eeq
in terms of the angles $\mu$ and $\phi$, with $\mu_1=\mu$, $\mu_2=\mu\cos\alpha-\sqrt{1-\mu^2}\sin\alpha\cos\phi$, for $\cos\alpha\equiv\hk_1\cdot\hk_2$. In practice, we must consider also the impact of the fiducial cosmology used to convert angles and redshifts into Cartesian coordinates: this modifies \eqref{eq: Bk-mon} to
\beq\label{eq: Bk-mon-AP}
    B_0(k_1,k_2,k_3) &=& \frac{1}{\alpha_\parallel^2\alpha_\perp^4}\int_{-1}^{1}\frac{d\mu}{2}\int_0^{2\pi}\frac{d\phi}{2\pi}\\\nonumber
    &&\qquad\,\times\,B(q_1[k_1,\mu_1],q_2[k_2,\mu_2],q_3[k_3,\mu_3],\nu_1[\mu],\nu_2[\mu_2(\mu,\phi)]) ,
\eeq
with $q=k\sqrt{\mu^2/\alpha_\parallel^2+(1-\mu^2)/\alpha_\perp^2}$ and $\nu=\mu/\left[\alpha_\parallel\sqrt{\mu^2/\alpha_\parallel^2+(1-\mu^2)/\alpha_\perp^2}\right]$ for conventional rescaling parameters $\alpha_\parallel$, $\alpha_\perp$ depending on cosmological parameters~\citep{Ivanov:2021kcd}. The angular integrals can be performed numerically via Gauss-Legendre quadrature in practice. We further incorporate integration over finite $k$-bins as well as discreteness effects; both effects are discussed below.

\section{Practical Implementation}\label{sec: implementation}

\vskip 8 pt

\subsection{Loop Integrals}
We now discuss how to compute the one-loop bispectrum. The most difficult part of this is evaluating the loop integrals appearing in \eqref{eq: Bk-one-loop}: in this work, these are computed via the FFTLog procedure \citep{Simonovic:2017mhp}, the subtleties of which are described in Appendix \ref{app:fftlog}. In essence, the real-space computation proceeds by first writing the integration kernels (products of $Z_n$) as polynomials in $k_i^2$, $q^2$, and $|\vk_i\pm \vq|^2$ (or their reciprocals). By expanding the linear (or IR-resummed) power spectrum as a sum over complex polynomials, \textit{i.e.}\ $P_L(k)\sim \sum c_m k^{\nu+i\eta_m}$ for frequency $\eta_m$ and FFTLog `bias' $\nu$, the various terms in \eqref{eq: Bk-one-loop} take the form (using $B_{222}$ as an example)
\beq\label{eq: fft-log-master}
    \sum_{m_1m_2m_3}c_{m_1}c_{m_2}c_{m_3}\int_{\vq}\frac{1}{q^{2\nu_1}|\vk_1-\vq|^{2\nu_2}|\vk_2+\vq|^{2\nu_3}}
\eeq
for some complex $\nu_i$. The integral can be evaluated using techniques borrowed from quantum field theory, and reduces the calculation to a tensor multiplication, noting that all cosmological information is encoded within $c_m$. In redshift space, the appearance of angles, $\hq\cdot\hn$ inside the integral make this more challenging; however, it can be evaluated using similar tricks to the one-loop power spectrum \citep[cf.][]{Chudaykin:2020aoj}), as discussed in Appendix \ref{app:fftlog}.

Following the above tricks, the bispectrum takes the following schematic form, again taking $B_{222}$ as an example:
\beq\label{eq: b222-expan}
    B_{222}(k_1,k_2,k_3,\mu_1,\mu_2) \sim \sum_i\bb{\theta_i}\sum_{0\leq j+k\leq 12}\mu^j\,\chi^k\, B^{(i,j,k)}_{222}(k_1,x,y),
\eeq
where the $i$ index runs over all combinations of bias and $f(z)$, denoted $\theta_i$.\footnote{Examples of these include $b_1^3, \gamma_2b_1f^2,f^6$ \textit{et cetera.}} Additionally, we have expanded in terms of the redshift-space angles $\{\mu,\chi\equiv\sqrt{1-\mu^2}\cos\phi\}$ (of which there are 47 non-trivial combinations); these are related to the $\mu_i$ angles via:
\beq
    \mu_1\equiv \mu, \quad \mu_2 \equiv \mu\cos\alpha-\chi\sin\alpha, \quad k_3\mu_3 = -k_1\mu_1-k_2\mu_2 .
\eeq
We adopt this basis rather than the more familiar choice of $\{\mu_1,\mu_2\}$, since it avoids pathologies for flattened triangles (whence $k_1\approx k_2+k_3$, and $\mu_1\approx -\mu_2$). The underlying shapes, $B^{(i,j,k)}$, appearing in \eqref{eq: b222-expan} are independent of both redshift-space angles and bias parameters, and depend only on the form of the linear power spectrum, $k_1$, $x = k_3^2/k_1^2$ and $y=k_2^2/k_1^2$, assuming $k_1\geq k_2\geq k_3$. 

Two options arise for using the bispectrum templates $B^{(i,j,k)}$ in Monte Carlo Markov Chain (MCMC) analyses: (a) they may be computed once for a fixed linear power spectrum, (b) they may be computed as a tensor multiplication (cf.\,\ref{eq: fft-log-master}) at each step in the MCMC chain, feeding in the relevant linear power spectrum (and thus $c_m$ coefficients) each iteration. Whilst (b) is the approach usually adopted for the one-loop power spectrum, we will here adopt (a) for the one-loop bispectrum. This has the effect of fixing cosmology in the bispectrum loops (except for $\sigma_8$, which acts as a global rescaling, modulo a small effect concerning the IR resummation amplitude, which we ignore in this work), and is chosen on computational grounds, since the size of the necessary FFTLog matrices becomes very large.\footnote{To see this, note that the matrix in \eqref{eq: fft-log-master} has size $N_{\rm freq}^3$, for $N_{\rm freq}$ FFTLog frequencies. Taking $N_{\rm freq}=64$, with $47$ angular combinations, $\mathcal{O}(50)$ bias parameter combinations, and computing the matrix for $10$ choices of each of $x$ and $y$ (noting that $k$ scales out), we find $\sim 5\times 10^{10}$ elements, or $\sim 50\,\mathrm{GB}$ in (complex) single precision.} Explicitly, we compute the bispectrum templates, $B^{(i,j,k)}$, for a grid of values of $\{x, y, k_1\}$ (treating flattened triangles with $\sqrt{x}+\sqrt{y}=1$ separately to avoid divergences), then use these to construct a three-dimensional linear interpolator for each shape. The resulting bispectra have been compared to results from explicit (and computationally intensive) numerical integration for a range of values of bias and triangle shapes and found to be in excellent agreement. Full details of the above steps are given in Appendix \ref{app:fftlog-imp}. We additionally publicly release all our analysis code: this can be found at \href{https://github.com/oliverphilcox/OneLoopBispectrum}{GitHub.com/OliverPhilcox/OneLoopBispectrum}.

\subsection{Bin Integration}
To robustly compare theory and data, we must integrate the model across some set of bins. Following \citep{Ivanov:2021kcd}, this is achieved via the integral:
\beq\label{eq: bk-binning}
    \hat B_{0,\rm int} = \frac{V^2}{(2\pi)^6N_{123}}\prod_{i=1}^3\left(\int_{k_i-\Delta k/2}^{k_i+\Delta k/2}dq_i\,q_i\right)B_0(q_1,q_2,q_3),
\eeq
where $B_0$ is the bispectrum monopole of \eqref{eq: Bk-mon-AP} and $N_{123} = 8\pi k_1k_2k_3(\Delta k)^3V^2/(2\pi)^6$ for bin center $(k_1,k_2,k_3)$ and width $\Delta k$. As in \citep{Ivanov:2021kcd}, this is strictly exact only in the narrow-bin limit, and can be corrected by ``discreteness weights'' as in the former work. In practice, we compute the set of bispectrum templates $B^{(i,j,k)}(k_1,x,y)$ for a range of values of $k_1,x,y$ (see Appendix \ref{app:fftlog-imp}) then perform the bin-averaging by linearly interpolating these values, dropping any triangles that do not satisfy the triangle conditions $|q_1-q_2|\leq q_3\leq q_1+q_2$. The integration is performed using Gauss-Legendre quadrature, as for the angular integrals. Finally, we note that we can perform bin integration either within the MCMC chains or as a pre-processing step (allowing us to use bin-averaged templates in the later analysis). We use the latter option for the purposes of this paper.

\subsection{Free Parameters}\label{subsec: free-params}
Our full model for the one-loop galaxy power spectrum and bispectrum depends on the following 44 free parameters (\textit{i.e.}\ Wilson coefficients):
\beq
    &&\,\,\,\,\,\{\textcolor{purple}{b_1},\textcolor{purple}{b_2},
    \textcolor{purple}{b_{\mathcal{G}_2}},b_3,\gamma_2^\times,\gamma_3,\textcolor{purple}{
   b_{\Gamma_3}
    },\gamma_{21}^\times,\gamma_{211},\gamma_{22},\gamma_{31}\}\\\nonumber
    &&\times\{\textcolor{blue}{c_0},\textcolor{blue}{c_2},\textcolor{blue}{c_4},\textcolor{blue}{\tilde c},\beta_{B,a},\beta_{B,b},\beta_{B,c},\beta_{B,d},\beta_{B,e},C_i [i=1...9]\}\\\nonumber
    &&\times\{\textcolor{purple}{P_{\rm shot}}, \textcolor{blue}{a_0}, 
    \textcolor{blue}{a_2},
    B_{\rm shot},
    A_{\rm shot,0},A_{\rm shot,1},S_i [i=0...7]\} ,
\eeq
where parameters appearing only in the power spectrum (following the definitions of \citep{Ivanov:2021kcd}), only in the bispectrum, and in both spectra, are shown in blue, black and purple respectively. The three lines give bias parameters, UV counterterms, and stochasticity parameters respectively. 
Note that here we switch to the power spectrum biases $b_{\mathcal{G}_2}$ and $b_{\Gamma_3}$
instead of $\gamma_2$ and $\gamma_{21}$ to ease the comparison
with previous works~\cite{Ivanov:2021fbu,Ivanov:2021kcd}; these are related via \eqref{eq: bias-relation}.
Whilst performing an MCMC analysis in this high-dimensional space may seem a formidable task, we note that all parameters except $\{b_1,b_2,b_{\mathcal{G}_2}\}$ enter the theory model linearly, and can thus be analytically marginalized, following \citep{Philcox:2020zyp}. This is exact, and will be applied to all analyses presented in this work, significantly reducing computational cost. Since the parameter $b_{\Gamma_3}$ is of physical interest in power spectrum analyses, we opt to marginalize over this explicitly, alongside the quadratic biases.

For the purposes of the analytic marginalization, 
we assume the following priors on the bispectrum 
nuisance parameters: all means are zeros, 
and the expectation values given by 10 
for all bias parameters, 10 for all real-space counterterms and one-loop stochastic contributions,
20 for redshift-space counterterms (in order to account for 
enhancements caused by short-scale non-linear 
redshift-space distortions, known as fingers-of-God~\cite{Jackson:2008yv}), 
and 20 for redshift-space one-loop stochastic contributions.
For the tree-level stochastic counterterms, following \citep{Ivanov:2021kcd}
we assume standard deviations of 5 for the dimensionless $B_{\rm stoch}$, $P_{\rm shot}$, and $A_{\rm shot}$ parameters. The power spectrum nuisance priors match \cite{Ivanov:2021fbu,Ivanov:2021kcd}.
Note that our nuisance parameters are normalized in such a way 
that their physical values are expected to be $\mathcal{O}(1)$
numbers from the naturalness arguments. In this sense our physically-motivated choice of nuisance parameter is conservative, as we allow them to be as large as $\mathcal{O}(10)$.

\subsection{Numerical Results}

Before proceeding to use the one-loop bispectra to perform parameter inference, we first consider the form of the spectra themselves. Plotting the bispectrum is a challenge itself: the monopole exists in the three-dimensional simplex of $\{k_1,k_2,k_3\}$, and we have contributions from a wide variety of nuisance parameter combinations. For the purpose of visualization, we will fix the bias parameters to simple local-in-Lagrangian space predictions, based on \citep{Eggemeier:2020umu}: assuming the bias to be described only by linear and quadratic terms $b_1^L, b_2^L$ in Lagrangian space, this gives 
\beq
    &&\left\{b_1 = b_1^L + 1,\,b_2 = b_2^L,\,b_3 = -3b_2,\,\gamma_2 = -\frac{2}{7}b_1^L,\,\gamma_3 = -\frac{1}{9}b_1^L - \gamma_2,\right.\\\nonumber
    &&\quad \left.\gamma_{21} = \frac{2}{21}b_1^L + \frac{6}{7}\gamma_2,\,\gamma_2^\times = -\frac{2}{7}b_2,\,\gamma_{22} = -\frac{6}{539}b_1^L-\frac{9}{49}\gamma_2\right.\\\nonumber
    &&\quad \left. \gamma_{21}^\times = \frac{2}{21}b_2 + \frac{2}{7}\gamma_2^\times,\,\gamma_{31} = -\frac{4}{11}b_1^L - 6\gamma_2,\,\gamma_{211} = \frac{5}{77}b_1^L + \frac{15}{14}\gamma_2 + \gamma_{21}-\frac{9}{7}\gamma_3\right\}.
\eeq

\begin{figure}
    \centering
    \includegraphics[width=\textwidth]{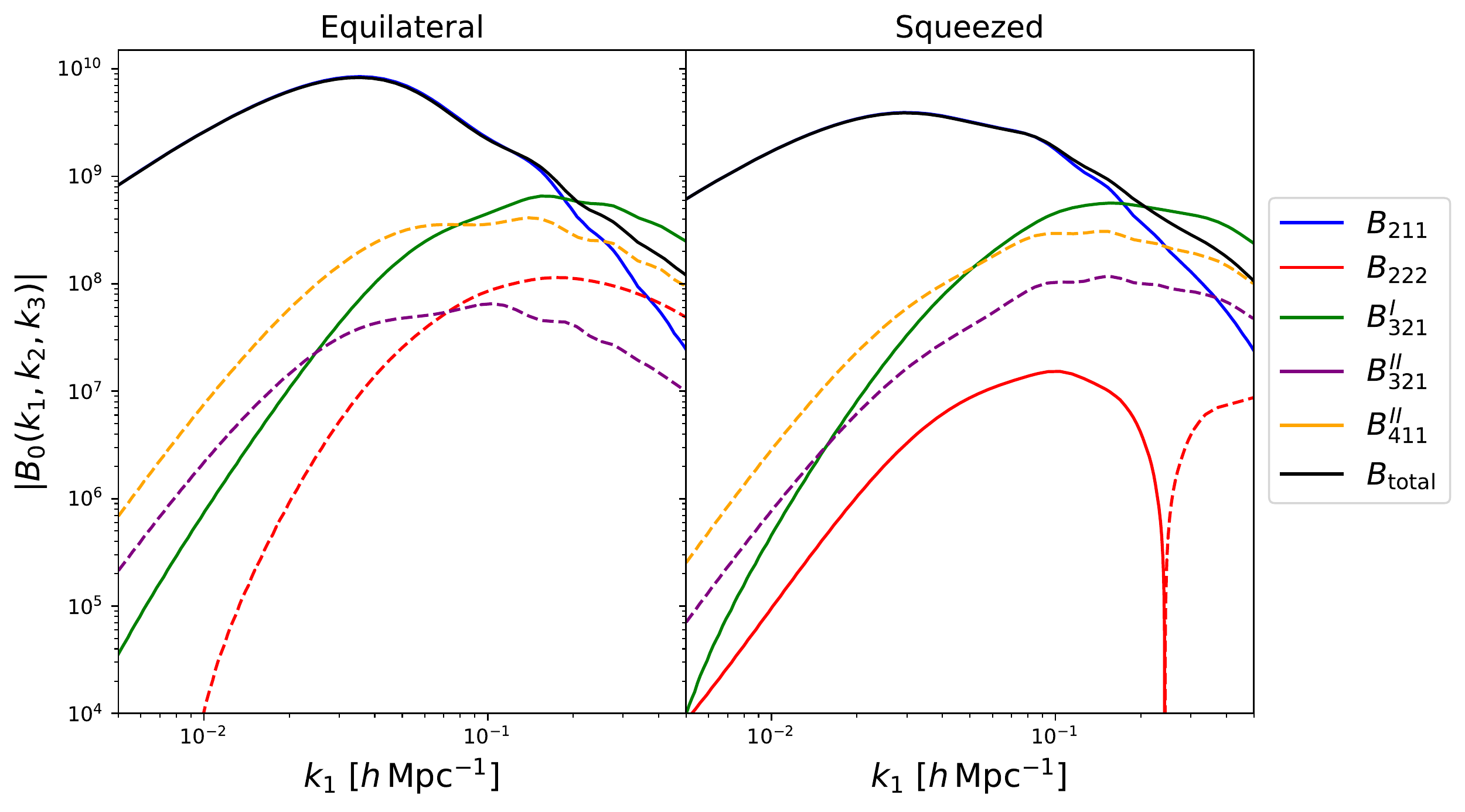}
    \caption{Contributions to the one-loop galaxy bispectrum monopole in redshift-space. The blue and black curves show the tree-level and one-loop bispectra respectively, with individual one-loop contributions shown in color. Dashed lines indicate negative contributions, and we show results for two types of triangle: equilateral, with $k_1=k_2=k_3$, and squeezed, with $k_2=0.9k_1$ and $k_3=0.2k_1$. For illustration, we assume coevolution biases following \citep{Eggemeier:2020umu}, with Lagrangian biases $b_1^L = 1$, $b_2^L = 0.3$ and a growth factor $f(z) = 0.7$. We do not include the contributions from stochasticity or counterterms in this plot, but note that all bias operators have been renormalized.}
    \label{fig: bk-plot}
\end{figure}

In Fig.\,\ref{fig: bk-plot} we plot the deterministic (Eulerian PT) bispectrum contributions assuming the above bias relations with $b_1^L = 1$, $b_2^L = 0.3$ and $f(z) = 0.7$, as well as distortion parameters $\alpha_\parallel=\alpha_\perp = 1$ and the best-fit PT Challenge input power spectrum (cf.\,\S\ref{sec: data}). For both equilateral and squeezed triangles we observe a similar form: the one-loop corrections are suppressed on large scales (by $k/k_{\rm NL}$) but become large as $k$ increases, with the $B_{321}^I$ piece exceeding tree-level theory by $k\sim 0.1\hMpc$. We find significant cancellation between the various one-loop components (which all depend on the same biases),
which is expected from 
the IR cancellation 
of loop integrals.
Note that the high-$k$ behavior is further modified by the counterterms (scaling as $k^2P_L^2(k)$) and stochasticity (scaling as $k^0$ and $P_L(k)$ at leading-order). The individual shapes of the bispectrum components are generally non-trivial, with oscillatory signatures seen in $B_{411}$ and, to a lesser extent, $B_{321}^{I,II}$. The smooth nature of $B_{222}$ (expected since the three power spectra are all inside the $\vq$ integral) implies that a smaller number of FFTLog frequencies can likely be used in its computation, which may expedite the template computation, and suggests that this has only weak cosmology dependence. From the deterministic contributions alone, it is clear that the one-loop bispectrum is a significant fraction of $B_{\rm tree}$ for all $k\gtrsim 0.1\hMpc$, and thus its inclusion is necessary if we wish to model the bispectrum beyond the softest modes.

\section{Data and Analysis Details}\label{sec: data}

The dataset used in this paper is the PT Challenge suite~\cite{Nishimichi:2020tvu}, comprising high-resolution $N$-body simulations at $z=0.61$ with a total volume of 566\,$h^{-3}\mathrm{Gpc}^3$. Galaxies are allocated via a BOSS-like halo occupation prescription, and various summary statistics computed using a fiducial cosmology with $\Omega_m = 0.3$. In all our analyses, we use the redshift-space power spectrum multipoles, $P_\ell(k)$, and the real-space power spectrum proxy $Q_0$, both of which were studied in detail in~\cite{Ivanov:2021fbu}. In this work, we additionally add the bispectra in both real- and redshift-space, with the comparison allowing us to assess the relative importance 
of redshift-space distortions in the one-loop bispectrum.

The relevant bispectra are computed as described in \cite{Ivanov:2021kcd}, which studied the tree-level bispectrum 
likelihood. We bin the bispectrum data in wavenumber bins of width $\Delta k = 0.01\hMpc$, and use only triangles whose bin centers satisfy momentum 
conservation.\footnote{As shown in \cite{Ivanov:2021kcd}, the remaining triangles do not carry significant cosmological information.} For $\kmax=0.15,~0.2,~0.3$, we find a total of 372, 825, and 2600 independent triangle configurations, respectively, and note that, unlike \cite{Ivanov:2021kcd}, we do not include the very first bin in the analysis, \textit{i.e.}\ we fix $k_{\rm min}=0.01\hMpc$ for the bispectrum.
This matches the analyses of the actual surveys like BOSS, where the very first bin is often affected by systematics including stellar contamination~\cite{Philcox:2021kcw,Alam:2016hwk}. 

Our theory model for the power spectrum matches that of \cite{Ivanov:2021fbu,Ivanov:2021kcd}, and we make use of the publicly available code \textsc{class-pt}~\cite{Chudaykin:2020aoj} to compute the power spectrum models.\footnote{\href{https://github.com/michalychforever/CLASS-PT}{GitHub.com/michalychforever/CLASS-PT}.}
Similarly, our theoretical model for the bispectrum is discussed in detail in Section \ref{sec: theory}, and implemented using the FFTLog prescription using \textsc{Mathematica} -- we refer the reader to Appendix~\ref{app:fftlog} for technical details.

An important part of the likelihoods are the covariance matrices, encoding both errors and correlations. As in previous works, we here adopt the Gaussian tree-level approximation for the analytic covariance matrices of power spectra and bispectra, neglecting any cross-correlation between the two statistics. For sufficiently large scales these assumptions are well justified \cite{Wadekar:2020hax,Ivanov:2021kcd,Oddo:2021iwq,Rizzo:2022lmh}; at smaller-scales, and in the presence of non-uniform survey geometry, a mock-based approach will probably be needed, such as in \citep{Philcox:2021kcw}, most likely in combination with some compression scheme \citep[e.g.,][]{Philcox:2020zyp}.

The mock galaxy clustering data from the PT Challenge simulations are analyzed within the Bayesian framework. Here, we perform a global MCMC analysis using the publicly available sampler
\textsc{Montepython}~\cite{Audren:2012vy,Brinckmann:2018cvx} varying the clustering amplitude $\sigma_8$, $f_{\rm NL}^{\rm equil}$ (the amplitude of equilateral primordial non-Gaussianity~\cite{Cabass:2022wjy}) 
and the EFT nuisance parameters. Since the true value of $\sigma_8$ in the simulations remains blinded, we will show results only for the fractional error on $\sigma_8$. As noted above, we will marginalize over all physical nuisance parameters given in \S\ref{subsec: free-params}. This is in contrast with some bispectrum studies that aim to fix certain nuisance parameters, such as asserting coevolution relations for Lagrangian biases \citep{Eggemeier:2021cam}. Indeed, for some particular purposes,
\textit{i.e.}\ fits of $\sigma_8$, it may be sufficient to keep fewer parameters in the fit. However, such approximations are unwarranted -- their validity can break down for other types of analyses. Therefore, we prefer to explicitly vary all physical nuisance
parameters in the fit.  By virtue of analytic marginalization~\cite{Philcox:2020zyp}, this is done at no computational cost.

\section{Results: Real-space}\label{sec: results-real}

We now present results from the above analyses, focusing first on the combination of the redshift-space power spectrum and real-space bispectrum. Though not quite matching observational setups (where the power spectrum and bispectrum are both observed in redshift-space), this analysis will allow us to understand the impact of redshift-space distortions.

To obtain the real space model, we set $f=0$ in all calculations and retain only EFT operators that do not depend on the LoS angles, giving a one-loop model fully equivalent to that used in~\cite{Eggemeier:2018qae,Eggemeier:2021cam}.
Our data vector contains the power spectrum multipoles, the real-space analog and the bispectrum monopole, \textit{i.e.}\ $[P_\ell(k),Q_0(k),B(k_1,k_2,k_3)]$ and we restrict to the $z=0.61$ snapshot of the PT Challenge simulations. In most analyses we use $P_\ell(k)$ up to $\kmax=0.16\hMpc$, and $Q_0$ in the range $0.16\hMpc\leq k<0.40\hMpc$, as validated in~\cite{Nishimichi:2020tvu,Ivanov:2021fbu,Ivanov:2021kcd}. We explore the impact of varying the bispectrum $k_{\rm max}$ below.

\subsection{Clustering amplitude and bias parameters}

\begin{figure}
\centering
\includegraphics[width=0.99\textwidth]{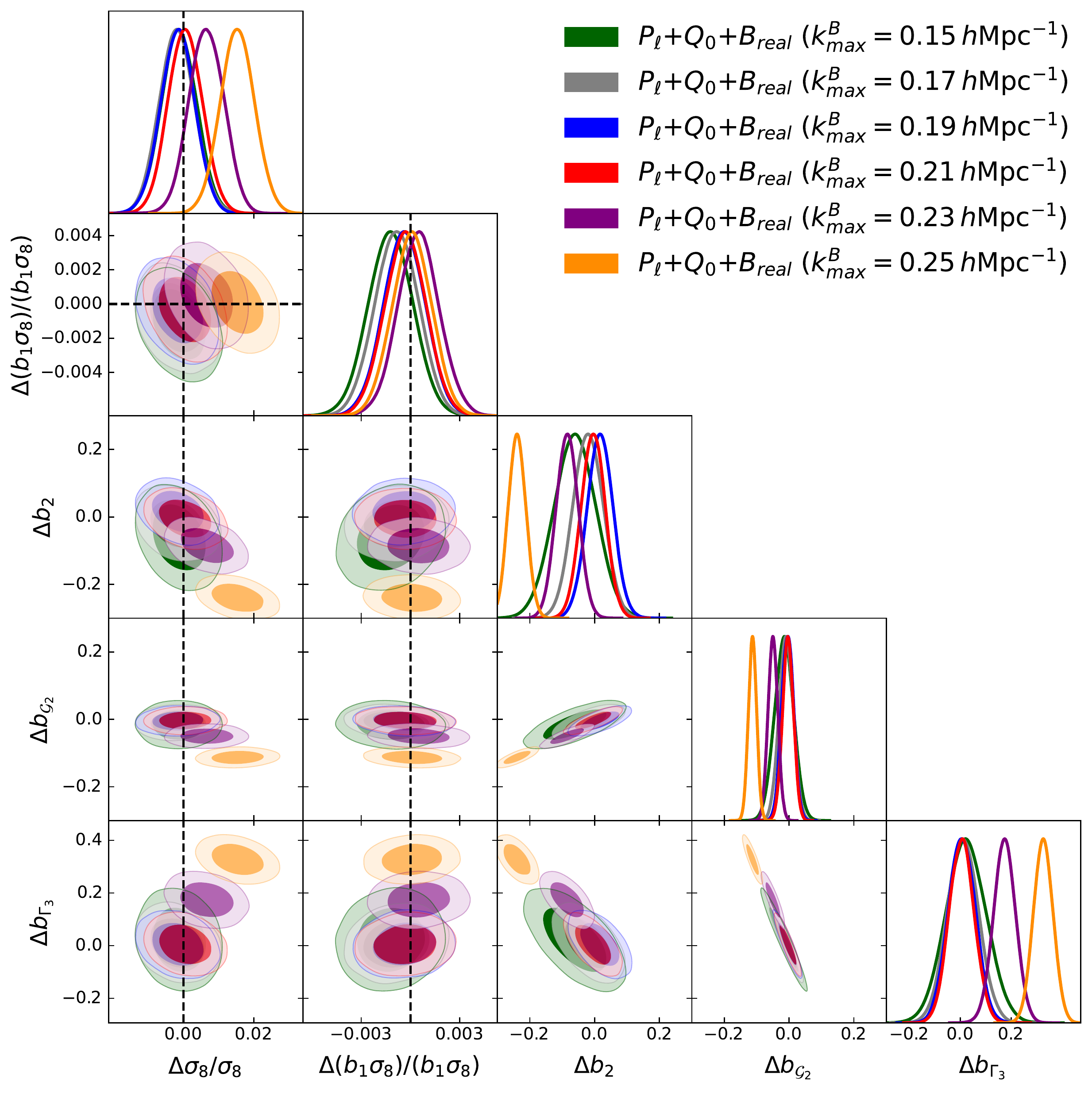}
\caption{
Posterior distributions for the clustering amplitude, $\sigma_8$, and certain nuisance parameters extracted from MCMC analyses of the power spectrum multipoles and the one-loop real-space bispectrum. The power spectrum likelihood is the same for all cases, whilst we vary $\kmax^B$ for the bispectrum, as indicated in the caption. Corresponding marginalized parameter contours for $\kmax=0.21\hMpc$ are given in Tab.\,\ref{tab:reals}.}
\label{fig:reals}
\end{figure}

We first focus on measuring the mass clustering amplitude $\sigma_8$
and leading galaxy bias parameters $\{b_1,b_2,b_{\mathcal{G}_2},b_{\Gamma_3}\}$. These appear both in the one-loop power spectrum and bispectrum models, and hence can be tightly constrained by the data. Unlike $\sigma_8$, the true values of bias parameters in the simulations are unknown. As such, we take their best-fit values at a certain $\kmax$
(where the one-loop model can be trusted) as a proxy for their true values. This $\kmax$ is measured as in \cite{Nishimichi:2020tvu} (see also~\cite{Chudaykin:2020hbf}) by determining at what scale cut posteriors for at least one parameter become biased w.r.t.~analyses with lower $\kmax$. We take best-fit values of bias parameters 
at the last stable $\kmax$ as ground truth, $p_{\rm true}$. Following this, our parameter measurements are quoted as $\Delta p = p - p_{\rm true}$, to avoid unblinding the results.

We fit the real-space bispectrum data for the following choices of scale-cut: 
\[
\kmax^B/(\hMpc)=0.15,~0.17,~0.19,~0.21,~0.23,~0.25\,.
\] 
The resulting posterior contours are shown in Fig.~\ref{fig:reals}, with one-dimensional marginalized limits on the amplitude and some nuisance parameters shown in Tab.\,\ref{tab:reals}. From the figure, we find that the posterior on $\sigma_8$ remains unbiased up to $\kmax^B=0.21\hMpc$, with a shift of $1-2\%$ observed for $\kmax^B=0.23\hMpc$ and $\kmax^B=0.25\hMpc$, which becomes significant relative to the PT Challenge error-bars.

However, at $\kmax^B\geq 0.23\hMpc$, we observe that the nuisance parameters become biased w.r.t.~measurements at low scale cuts, for example, we find a visible tension between the $b_2$ posterior at $\kmax^B=0.23\hMpc$ and $\kmax^B=0.19\hMpc$. In addition, we see that parameters the $b_{\mathcal{G}_2}$ and $b_{\Gamma_3}$ become biased.
The optimal values of these parameters scale along the degeneracy direction $b_{\mathcal{G}_2}+0.34b_{\Gamma_3}\approx \text{const}$, which closely matches the degeneracy combination
imposed by the power spectrum, $b_{\mathcal{G}_2}+0.4b_{\Gamma_3}$~\cite{Ivanov:2019hqk}. 

In contrast with the $\kmax^B\geq 0.21~\hMpc$ picture, the results at all choices of $\kmax^B\leq 0.21\hMpc$ are fully consistent, implying that $\kmax^B =  0.21\hMpc$ should be chosen as a baseline scale cut. This is somewhat larger than the one-loop power spectrum scale-cut of $k_{\rm max} = 0.16\hMpc$; whilst this might appear unusual, we note that the power spectrum contains significantly higher signal-to-noise, and is subject to redshift-space complexities, both of which decrease $k_{\rm max}^P$. We use best-fit values of nuisance parameters from the baseline $P_\ell+Q_0+B(k_{\rm max}^B=0.21\hMpc)$ analysis as ground truth values in the below.

\begin{table}[!htb]
    \begin{minipage}{0.5\linewidth}
\vspace{0.1cm}
\begin{tabular}{|l|c|c|}
\hline
\multicolumn{2}{|c|}{$P_\ell$}
\\ \hline
Parameter & 68\% limits \\ \hline
\hline

$\Delta \sigma_8/\sigma_8  $ & $0.0080\pm 0.0064          $\\

$\Delta (b_1\sigma_8)/(b_1\sigma_8)    $ & $-0.0063^{+0.0042}_{-0.0038}$\\

$\Delta b_2                $ & $-0.04^{+0.44}_{-0.56}     $\\

$\Delta b_{\mathcal{G}_2}  $ & $0.08^{+0.25}_{-0.34}      $\\
$\Delta b_{\Gamma_3}       $ & --- \\
\hline
\end{tabular}
\end{minipage}%
    \begin{minipage}{.5\linewidth}
\begin{tabular}{|l|c|c|}
\hline
\multicolumn{2}{|c|}{\text{$P_\ell$+$Q_0$+$B_{\rm real}$}}
\\ \hline
Parameter & 68\% limits \\ \hline

\hline

$\Delta \sigma_8/\sigma_8  $ & $0.0005\pm 0.0049          $\\

$\Delta (b_1\sigma_8)/(b_1\sigma_8)    $ & $-0.0003\pm 0.0013         $\\

$\Delta b_2                $ & $-0.005\pm 0.037           $\\

$\Delta b_{\mathcal{G}_2}  $ & $-0.003\pm 0.017           $\\

$\Delta b_{\Gamma_3}       $ & $0.005\pm 0.050            $\\

\hline 
\end{tabular}
\end{minipage}%

 \caption{One-dimensional marginalized contraints on low-order bias parameters and the clustering amplitude $\sigma_8$ extracted from the PT Challenge data-set. We display results obtained using only the power spectrum multipoles $P_\ell$ (left panel, cf.\,\citep{Nishimichi:2020tvu}), and those including the power spectrum, $Q_0$ and the one-loop real-space bispectrum likelihood with $\kmax^B=0.21\hMpc$ (right panel). The one-loop bispectrum is the main new feature of this work. Most parameters are normalized to their true values, to avoid unblinding the simulation. In real-space, the addition of the bispectrum significantly tightens posteriors on bias parameters (by at least an order of magnitude), and gives $\approx 20\%$ improvement on $\sigma_8$. Further details are given in the main text, with corresponding results for the redshift-space bispectrum shown in Tab.\,\ref{tab:rsds}.
\label{tab:reals}
}
\end{table}

It is instructive to compare the parameter constraints extracted using the one-loop bispectrum to those from the power spectrum multipoles alone, \textit{i.e.}\ $P_\ell(k)$ at the baseline $\kmax=0.16\hMpc$. 
These are shown in the left panel of Tab.\,\ref{tab:reals}. We find an improvement of $31\%$ in $\sigma_8$, whilst the error-bars on bias parameters tighten by an order of magnitude in some cases. Despite the noticeable increase in signal-to-noise of the data-set, we find a modest improvement in $\sigma_8$: this is linked to the proliferation of bias, counterterm, and stochasticity parameters needed to describe the one-loop bispectrum in an unbiased manner.

It is also useful to compare our one-loop bispectrum results with those from the tree-level bispectrum. We cannot directly use results from \cite{Ivanov:2021kcd} since the former work also varied other cosmological parameters such as $H_0$ and $\Omega_m$. To obtain a cleaner comparison, we repeat the tree-level analysis of \cite{Ivanov:2021kcd} with the same analysis settings as here, using $\kmax^B=0.08\hMpc$ for the tree-level bispectrum. We find $\Delta \sigma_8/\sigma_8 = 0.002 \pm 0.0053$, \textit{i.e.}\ an $21\%$ improvement over the power spectrum only result. Comparing this the present analysis, we see that the addition of the one-loop bispectrum likelihood yields an extra $10\%$ improvement over the tree-level bispectrum likelihood.

Finally, we compare our results with those of \cite{Eggemeier:2021cam}. Unlike our work, \cite{Eggemeier:2021cam} used the power spectrum of halos and galaxies in real space, leading to the notorious $b_1-\sigma_8$ degeneracy being largely unbroken. 
This explains why our results on $\sigma_8$ are much better -- most of the constraining power comes from redshift-space distortions omitted in \cite{Eggemeier:2021cam}. Despite this difference, our analysis does confirm a general trend pointed out in \cite{Eggemeier:2021cam} -- the returns from the one-loop bispectrum are limited by the large number of nuisance parameters. As such, it will be important to obtain better priors on them in the future, for example using hydrodynamical simulations.

\subsection{Primordial non-Gaussianity}\label{subsec: png}

It is interesting to study to what extent the one-loop bispectrum model 
can help improve constraints on primordial non-Gaussianity (PNG), following constraints from the tree-level bispectrum in \citep{Cabass:2022wjy,Cabass:2022ymb}. We consider here the case of equilateral PNG, which
induces the following three-point correlation of the linear density field (see~\cite{Cabass:2022wjy} for further details),
\be
\begin{split}
& \langle \delta^{(1)}_{\k_1}\delta^{(1)}_{\k_2}\delta^{(1)}_{\k_3}\rangle =(2\pi)^3
\delta_D^{(3)}(\k_{123})B_{\rm equil}(k_1,k_2,k_3)\,,\\
& B_{\rm equil} = \prod_{a=1}^3 T(k_a)\left[ 
\frac{18}{5}f^{\rm equil}_{\rm NL}\Delta^4_\zeta\frac{{\cal S}_{\rm equil}(k_1,k_2,k_3)}{k_1^2k_2^2k_3^2}
\right]\,,\\
& {\cal S}_{\rm equil}(k_1,k_2,k_3) = \bigg(\frac{k_1}{k_2} + \text{$5$ perms.}\bigg)   - \bigg(\frac{k_1^2}{k_2k_3} + \text{$2$ perms.}\bigg) - 2\,\,,
\end{split}
\ee 
where $\zeta$ is the primordial curvature 
fluctuation with dimensionless amplitude $\Delta_\zeta$, and we have introduced 
the transfer function 
$T(k)=(P_{11}(k)/P_\zeta(k))^{1/2}$.

Non-Gaussianity in the initial conditions 
generates three main effects~\cite{Cabass:2022wjy}: 
(1) an additional contribution $B_{111}$ to the tree-level bispectrum,
(2) an extra one-loop power spectrum correction $P_{12}$,
and (3) further contributions in the galaxy bias expansion, which 
modifies the tree-level expressions by introducing the so-called ``scale-dependent''
bias. 
The latter stems from the following expression,
\be
\delta_g \subset b_\zeta \nabla^2 \zeta + ... \;,
\ee
where ``$...$'' denote non-linear PNG corrections which can be ignored for the purposes of this paper. The term $\nabla^2 \zeta$
generates tree-level ``scale-dependent'' 
bias corrections to the power spectrum. Note that these corrections 
are suppressed in the equilateral case w.r.t.~the case of local PNG,
where the scale-dependent bias is a leading effect
on the galaxy power spectrum, and thus the power spectrum dominates the constraining power on $f_{\rm NL}^{\rm loc}$.

As shown in \cite{Cabass:2022wjy}, for the one-loop power spectrum and tree-level bispectrum, we must include all three of the $P_{12}$, $B_{111}$ and $b_\zeta$-related terms in our model. In this paper, we consider the one-loop bispectrum, which technically requires additional non-linear $f^{\rm equil}_{\rm NL}$ corrections to the galaxy bispectrum, such as $B^{(I)}_{113},B^{(II)}_{113},B^{(I)}_{122},B^{(II)}_{113}$~\cite{Assassi:2015jqa,Desjacques:2016bnm}. However, given that $f_{\rm NL}^{\rm eq}\Delta_\zeta$ is small, these will be suppressed, thus we leave their systematic calculation 
for future work, focussing only on the leading terms, similar to~\cite{Cabass:2022wjy,DAmico:2022gki}. For $f_{\rm NL}^{\rm equil}\lesssim 500$, the next-to-leading order contributions are subdominant to two-loop matter corrections for $k\lesssim 0.2\hMpc$.

Concerning scale-cuts, we find that use of the $P_\ell$ and $Q_0$ statistics at high $\kmax$ can lead to biases in the recovered values of $f_{\rm NL}^{\rm equil}$. This is consistent with the estimates of~\cite{Cabass:2022wjy}, which showed that 
the two-loop corrections can actually be larger than the non-Gaussian $P_{12}$ contribution at small scales. Thus, we choose $\kmax=0.2\hMpc$ for the $Q_0$ statistics 
and $\kmax=0.14\hMpc$ for $P_\ell$
in the PNG analysis of this section.
For $B_{\rm real}$ we use the baseline data cut $\kmax^B = 0.21\hMpc$, motivated by the discussion above.

To perform the analyses including PNG, we fit the parameter $f_{\rm NL}^{\rm equil}$ in addition to $\sigma_8$
and nuisance parameters. Since the PT challenge simulations were run using purely Gaussian initial conditions, we expect 
to find $f_{\rm NL}^{\rm equil}$ consistent with zero. Indeed, our nominal constraint on the amplitude of the equilateral shape is given by  
\be
\text{1-loop ($\kmax^B=0.21\hMpc$): }
\quad f_{\rm NL}^{\rm equil}= 72 \pm 80 ~(95\%~\text{CL}) \,. 
\ee 
We stress that these results are obtained without any external priors 
on $\sigma_8$ or the non-linear bias coefficients. This constraint can be compared with that obtained from the tree-level real-space bispectrum likelihood,
\be 
\text{tree ($\kmax^B=0.08\hMpc$): }\quad  f_{\rm NL}^{\rm equil}=53\pm 114~(95\%~\text{CL}) \,. 
\ee
At face value, our results imply that the addition of the one-loop bispectrum 
can improve constraints on $f_{\rm NL}^{\rm equil}$ by $\sim 30\%$.

\begin{figure}
\centering
\includegraphics[width=0.99\textwidth]{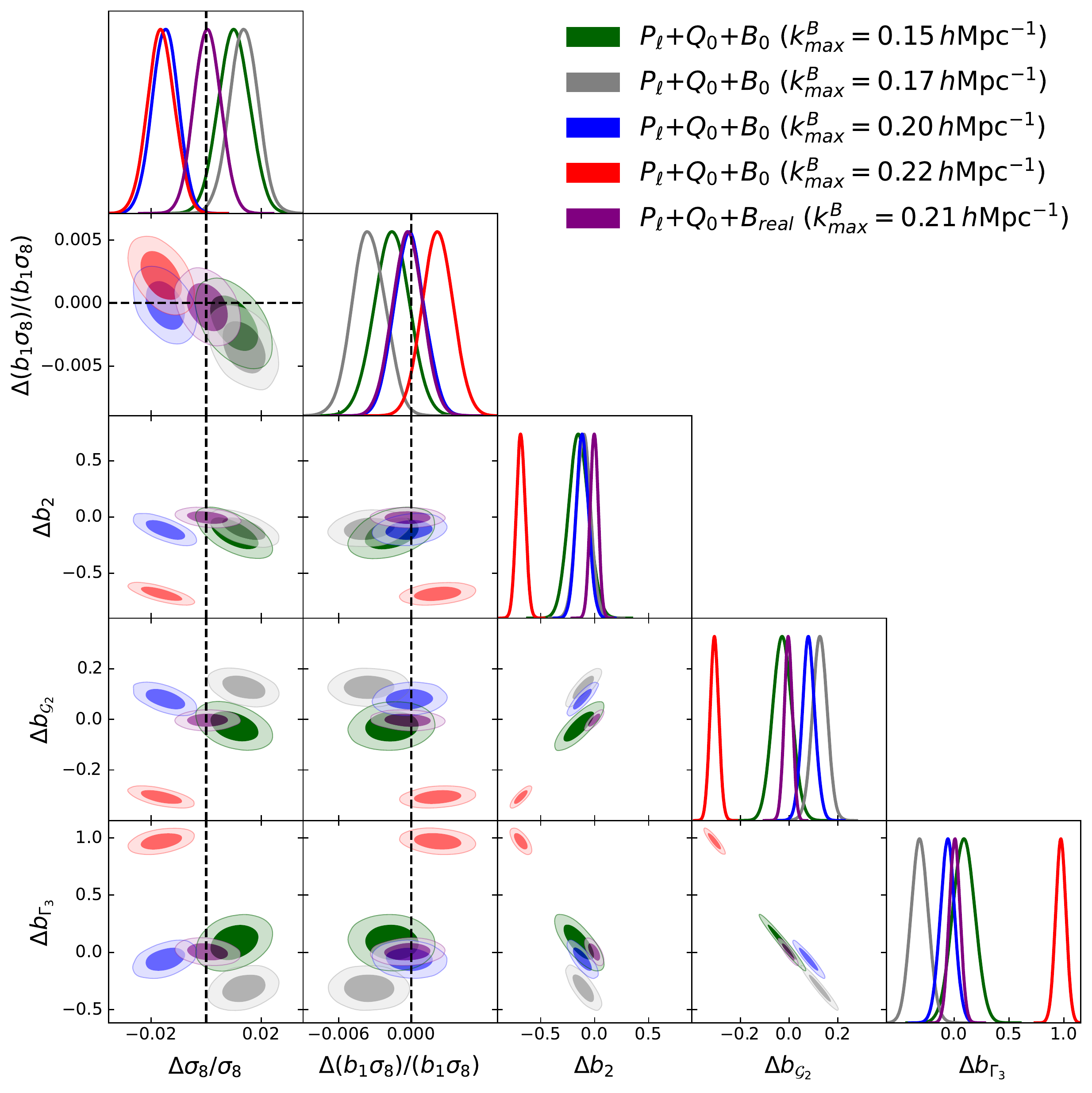}
\caption{
Posterior distributions of the clustering amplitude
and certain nuisance parameters obtained from MCMC analyses of 
the power spectra and one-loop redshift-space bispectrum monopole
$B_0$. The power spectrum likelihood is the same for all cases. 
We show results for different values of the bispectrum data cut $\kmax^B$, as indicated by the caption. This is analogous to Fig.\,\ref{fig:reals} (whose optimal constraint is shown by the purple curve), but utilizes the redshift-space bispectrum. Corresponding marginalized parameter constraints with $\kmax^B=0.15\hMpc$ are given in Tab.\,\ref{tab:rsds}. 
}
\label{fig:rsds}
\end{figure}

\section{Results: Redshift-Space}\label{sec: results-rsd}

In this section we present the analysis of the 
data-vector $[P_\ell(k),Q_0(k),B_0(k_1,k_2,k_3)]$,
where all statistics are in redshift space 
and include projection and coordinate-distortion effects. This set-up thus fully matches an analysis of a realistic
galaxy survey such as BOSS~\cite{Alam:2016hwk}. As for the power spectrum, we expect that the addition of redshift-space distortions (particularly the fingers-of-God effect~\cite{Jackson:2008yv}, hereafter FoG), will reduce the non-linear scale, thus it is likely that $\kmax^B$, and the constraining power of the bispectrum monopoole, will decrease.

\subsection{Clustering amplitude and bias parameters}

Let us discuss the recovery of the mass clustering amplitude $\sigma_8$ and 
bias parameters. As a point of comparison, we fix the fiducial bias parameters to those extracted from the real space analysis before with $\kmax^B=0.21\hMpc$. We fit the redshift-space bispectrum data for the following choices of scale-cut:
\beq
    \kmax^B/(\hMpc)=0.15,~0.17,~0.20,~0.22\,.
\eeq

\begin{table}[!htb]

    \begin{minipage}{.5\linewidth}
\begin{tabular}{|l|c|c|}
\hline
\multicolumn{2}{|c|}{\text{$P_\ell$+$Q_0$+$B^{\rm tree}_{0}$}}
\\ \hline
Parameter & 68\% limits \\ \hline

\hline

$\Delta \sigma_8/\sigma_8  $ & $0.0003\pm 0.0054          $\\

$\Delta (b_1\sigma_8)/(b_1\sigma_8)$ & $-0.0025\pm 0.0016         $\\

$\Delta b_2                $ & $-0.085\pm 0.082           $\\

$\Delta b_{\mathcal{G}_2}  $ & $0.034\pm 0.046            $\\

$\Delta b_{\Gamma_3}       $ & $-0.12\pm 0.12             $\\

\hline 
\end{tabular}
\end{minipage}%
    \begin{minipage}{.5\linewidth}
\begin{tabular}{|l|c|c|}
\hline
\multicolumn{2}{|c|}{\text{$P_\ell$+$Q_0$+$B^{\rm 1-loop}_{0}$}}
\\ \hline
Parameter & 68\% limits \\ \hline

\hline

$\Delta \sigma_8/\sigma_8  $ & $0.0101\pm 0.0057          $\\

$\Delta (b_1\sigma_8)/(b_1\sigma_8)$ & $-0.0016\pm 0.0015         $\\

$\Delta b_2                $ & $-0.146\pm 0.094           $\\

$\Delta b_{\mathcal{G}_2}  $ & $-0.028\pm 0.039           $\\

$\Delta b_{\Gamma_3}       $ & $0.09\pm 0.10              $\\

\hline 
\end{tabular}
\end{minipage}%
 \caption{One-dimensional marginalized constraints on low-order bias parameters and the clustering amplitude $\sigma_8$ 
extracted from the PT Challenge data-set. We show the fit from the combined likelihood including power spectrum multipoles, $Q_0$,
the tree-level 
redshift-space bispectrum at $\kmax^B=0.06\hMpc$ (left table)
and the 
one-loop redshift-space bispectrum at $\kmax^B=0.15\hMpc$ (right table). The parameters are normalized relative to their true values. Whilst we find significant enhancements in the bias parameter constraints compared to the power spectrum alone (cf.\,Tab.\,\ref{tab:reals}), the constraint on $\sigma_8$ does not improve appreciably.
\label{tab:rsds}
}
\end{table}

The resulting posterior contours are shown in Fig.~\ref{fig:rsds}, and we observe that all parameters of interest are unbiased at $\kmax^B=0.15\hMpc$. For comparison, we also show the baseline real space bispectrum results of \S\ref{sec: results-real}. 
Note that if we use $\kmax^B<0.15\hMpc$, the bispectrum data is not 
sufficient to constrain all the nuisance parameters entering 
the theory model. This gives rise to significant 
marginalization projection effects, which can be na\"ively interpreted 
as a bias in our model. We study these effects in Section~\ref{sec:proj}
and show that the measurements at $\kmax^B<0.15\hMpc$ 
are consistent with our baseline choice $\kmax^B=0.15\hMpc$
once projection effects are taken into account.

At $\kmax^B=0.17\hMpc$ the clustering amplitude $\sigma_8$, the rescaled linear bias $b_1\sigma_8$, $b_2$ and $b_{\Gamma_3}$ become biased w.r.t.~their optimal values coming from the real-space bispectrum analysis. These biases are accompanied with 
a significant increase in the $\chi^2$ statistics.
Thus, we conclude that the two-loop bispectrum corrections are not negligible at this scale. This is further supported by the bias growing with $\kmax^B$: in particular, at $\kmax^B=0.22\hMpc$ the bias on $\sigma_8$ reaches $2\%$, which is significant in the context of the PT Challenge simulation volume. In conclusion, we find that the one-loop galaxy bispectrum model in redshift space works well up to $\kmax^B=0.15\hMpc$
for the precision that corresponds to the total volume of the PT Challenge simulation (which we recall is significantly larger than current and forthcoming datasets). The optimal values of cosmological and bias parameters 
for this choice are presented in Tab.\,\ref{tab:rsds}. For comparison, we also show the results from the 
tree-level bispectrum analysis akin to~\cite{Ivanov:2021kcd}.\footnote{For this analysis we use $\kmax^B=0.06\hMpc$ and set the $c_1$ counterterm to zero for consistency with our one-loop model. This term served as a proxy for the one-loop corrections thus its inclusion is equivalent to partially accounting for one-loop information. To clearly compare the tree-level and one-loop results, we exclude it here.} 
That $\kmax^B$ is lower in redshift-space than real-space is no surprise: this indicates that the characteristic scale of FoG effects ($\sigma_{\rm FoG}$) is smaller than that of non-linearities ($k_{\rm NL}^{-1}$). For the power spectrum in redshift-space, higher-order counterterms were important to model FoG, scaling as $k_z^4$. An analogous set of nuisance parameters may be included here, but we caution that their number is large due to the higher dimensionality of the bispectrum.

Considering the marginalized posteriors directly, we find that the one-loop bispectrum likelihood (at $\kmax^B=0.15\hMpc$) 
yields only a $12\%$ improvement on $\sigma_8$ compared to the power spectrum alone, though the constraints on bias parameters (and thus astrophysics) improve markedly. Comparing this with the tree-level case, we see that the inclusion of one-loop corrections actually lead to a somewhat worse result than for tree-level bispectrum, which tightens the $\sigma_8$ constraint by $18\%$
for the analysis settings adopted in this work. This is consistent with previous studies considering the real-space bispectrum \citep{Eggemeier:2021cam}, and arises primarily due to the large number of nuisance parameters appearing in the one-loop calculation, especially in redshift-space. A similar situation takes place in the context of the one-loop redshift-space power spectrum, whose information content is limited by marginalization over nuisance parameters~\cite{Wadekar:2020hax}. We will discuss this issue in detail later. 

\section{Implications for the BOSS survey}\label{sec: boss}

\begin{figure}
\centering
\includegraphics[width=0.8\textwidth]{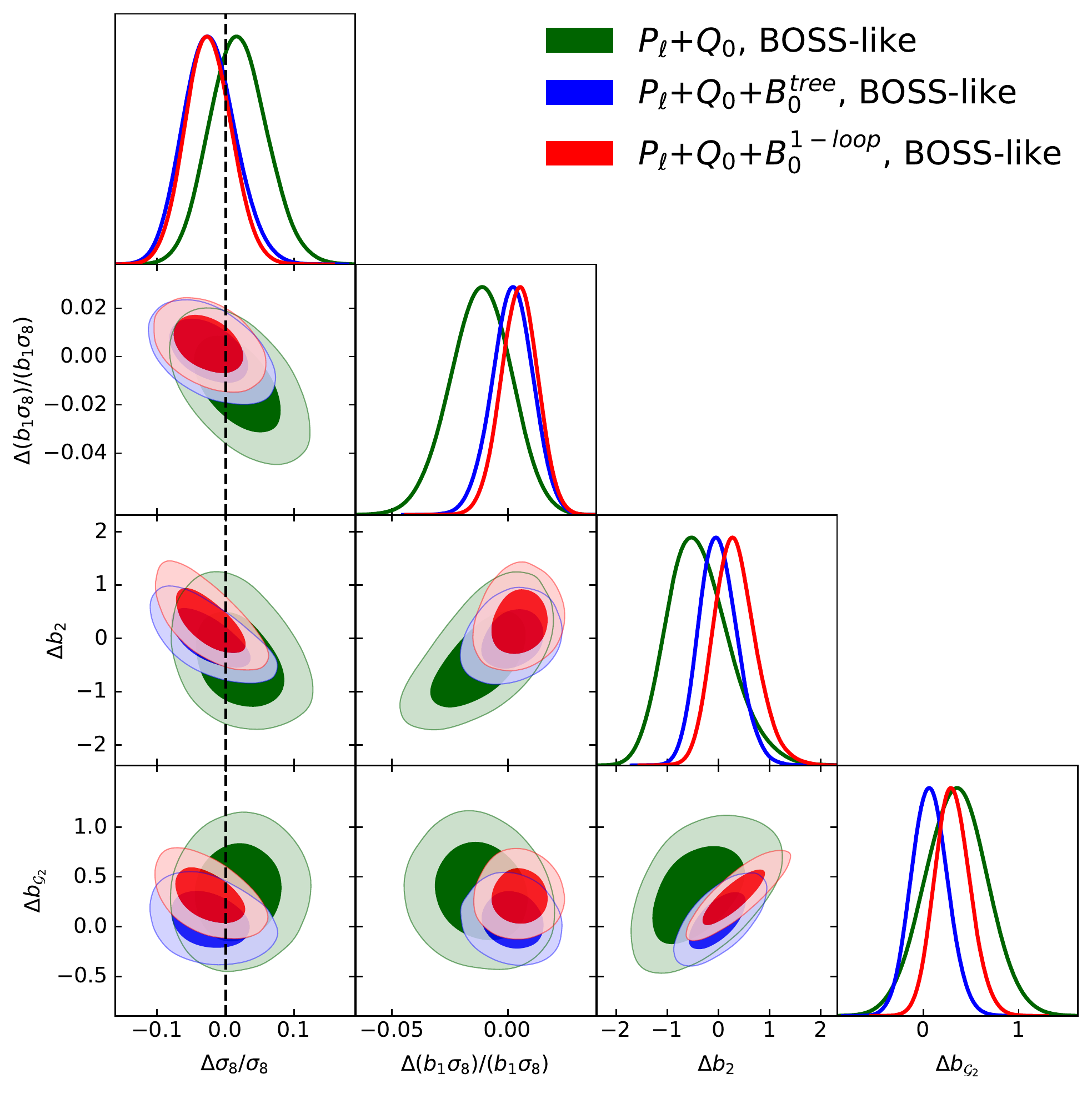}
\caption{
Posterior distributions of the clustering amplitude
and low-order nuisance parameters from MCMC analyses of 
the power spectrum data ($P_\ell+Q_0$, in green), 
and the combination of the power spectrum and redshift-space bispectrum monopole ($P_\ell+Q_0+B_0$), using tree-level (blue, from \citep{Ivanov:2021kcd}) and one-loop (red) theory. The covariance is rescaled to match the volume of the BOSS survey, and we assume $\kmax=0.2~\Mpch$ for both the one-loop 
power spectrum and bispectrum.} 
\label{fig:boss-like}
\end{figure}

In this section we estimate the potential performance of the one-loop bispectrum model applied to the BOSS survey data~\cite{Alam:2016hwk}, which is the largest publicly-available spectroscopic galaxy clustering dataset. This survey has significantly smaller volume than our mock simulation data, so one can expect that the analysis 
can be pushed to smaller scales~\cite{Chudaykin:2020ghx,Ivanov:2021kcd}. Indeed, the relevant parameter
in this problem is the ratio of the theory systematic bias in a certain parameter to the statistical error on that parameter.
For the BOSS volume the statistical errors are significantly larger than the PT Challenge simulation volume, due to a ratio of volumes of $\approx 100$. As the theoretical errors do not depend on the volume, the ratio between the theoretical
error and statistical errors thus becomes smaller, and hence any residual theoretical systematics becomes less sizable in relative terms.

To demonstrate this, we repeat the likelihood analysis above for the redshift-space data vector $[P_\ell,Q_0,B_0]$, but 
rescale the covariance to match the BOSS volume $V_{\rm BOSS}=V_{\rm PT~Challenge}/100\simeq 6$ $(h^{-1}\text{Gpc})^3$. We select $\kmax=0.20\hMpc$ for power spectrum multipoles $P_\ell$ and  $\kmax^B=0.20\hMpc$ for the bispectrum monopole $B_0$; significantly larger than that found in \S\ref{sec: results-rsd}. The results of this analysis are presented in Fig.~\ref{fig:boss-like} and in Tab.\,\ref{tab:boss}, showing also the $P_\ell+Q_0$ and 
$P_\ell+Q_0+B_0^{\rm tree}$ ($\kmax^B=0.08~\hMpc$ following~\cite{Ivanov:2021kcd})
analyses for comparison.

\begin{table}[!htb]
    \begin{minipage}{0.5\linewidth}
\vspace{0.1cm}
\begin{tabular}{|l|c|c|}
\hline
\multicolumn{2}{|c|}{$P_\ell+Q_0$, BOSS-like}
\\ \hline
Parameter & 68\% limits \\ \hline
\hline

$\Delta \sigma_8/\sigma_8  $ & $0.019\pm 0.042            $\\

$\Delta (b_1\sigma_8)/(b_1\sigma_8)$ & $-0.013\pm 0.013           $\\

$\Delta b_2                $ & $-0.37^{+0.50}_{-0.68}     $\\

$\Delta b_{\mathcal{G}_2}  $ & $0.37\pm 0.33              $\\

\hline
\end{tabular}
\end{minipage}%
    \begin{minipage}{.5\linewidth}
\begin{tabular}{|l|c|c|}
\hline
\multicolumn{2}{|c|}{\text{$P_\ell$+$Q_0$+ $B^{\rm tree}_{0}$, BOSS-like}}
\\ \hline
Parameter & 68\% limits \\ \hline

\hline

$\Delta \sigma_8/\sigma_8  $ & $-0.023^{+0.035}_{-0.040}  $\\

$\Delta (b_1\sigma_8)/(b_1\sigma_8)$ & $0.0021\pm 0.0089          $\\

$\Delta b_2                $ & $-0.02^{+0.34}_{-0.40}     $\\

$\Delta b_{\mathcal{G}_2}  $ & $0.07\pm 0.19              $\\

\hline
\end{tabular}
\end{minipage}%
\\
\centering{
    \begin{minipage}{.5\linewidth}
\begin{tabular}{|l|c|c|}
\hline
\multicolumn{2}{|c|}{\text{$P_\ell$+$Q_0$+ $B^{\rm 1-loop}_{0}$, BOSS-like}}
\\ \hline
Parameter & 68\% limits \\ \hline

\hline

$\Delta \sigma_8/\sigma_8  $ & $-0.025\pm 0.034           $\\

$\Delta (b_1\sigma_8)/(b_1\sigma_8)$ & $0.0051\pm 0.0080          $\\

$\Delta b_2                $ & $0.31^{+0.35}_{-0.45}      $\\

$\Delta b_{\mathcal{G}_2}  $ & $0.31^{+0.17}_{-0.19}      $\\

\hline
\end{tabular}
\end{minipage}
}
 \caption{One-dimensional marginalized constraints on low-order bias parameters and the clustering amplitude $\sigma_8$ extracted from the PT Challenge dataset, with covariance adjusted to match the volume of the BOSS survey. We show results for the $P_\ell+Q_0$ only analysis (top left), the tree-level $P_\ell+Q_0 +B_0$ likelihood (top right), and the one-loop $P_\ell+Q_0 +B_0$ likelihood (bottom). The inclusion of the bispectrum sharpens constraints on $\sigma_8$ by $\approx 24\%$, with some $\approx 10\%$ improvement arising from the addition of the one-loop contributions.
\label{tab:boss}
}
\end{table}

We see that in the context of BOSS, the addition of the one-loop bispectrum yields an $\approx 24\%$ improvement over the power spectrum-only result and an $\approx 10\%$ improvement over the tree-level bispectrum likelihood result.
However, this leads to a 
noticeable shift in nuisance parameters, with $b_{\mathcal{G}_2}$ approximately $1.7\sigma$ from its fiducial value. This could simply be a prior-volume effect however (since the effect of the priors becomes more important at lower simulation volume), especially given that $b_{\mathcal{G}_2}$ departs from its fiducial value at $1.1\sigma$ already for the power-spectrum alone. The tree-level 
bispectrum
analysis, however, 
results in an
unbiased recovery of all 
nuisance and cosmological 
parameters.

It is also instructive to study whether the one-loop bispectrum can improve constraints on equilateral PNG. Incorporating this parameter in the analysis as before (varying both $f_{\rm NL}^{\rm equil}$ and $\sigma_8$) we find $f_{\rm NL}^{\rm equil}=197\pm 350$. For the clustering amplitude we find $\Delta \sigma_8/\sigma_8 =-0.026 \pm 0.035$, with a slight $0.6\sigma$ shift w.r.t.~the ground truth. For comparison, we have also run an analysis using the tree-level bispectrum 
at $\kmax^B=0.08\hMpc$ instead of the one-loop bispectrum, and found  $f_{\rm NL}^{\rm equil}=420\pm 440$, 
$\Delta \sigma_8/\sigma_8 =-0.025 \pm 0.040$.
First, we see some bias in $f_{\rm NL}^{\rm equil}$, which can be attributed to prior volume effects and somewhat more optimistic data cuts for the power
spectrum that we use in our analysis here. 
Indeed, in \cite{Cabass:2022wjy} it was shown that the tree-level model yields unbiased results on $f_{\rm NL}^{\rm equil}$ for $\kmax^B=0.08\hMpc$ and $\kmax^{P_\ell}=0.17\hMpc$.
Second, we notice that the bound on $f_{\rm NL}^{\rm equil}$ 
in the one-loop case is $30\%$ better than that of the tree-level analysis.
The improvement is quite modest as a consequence of the fact that the one-loop model introduces many nuisance parameters, which cannot be constrained by the data. In our analysis we use highly conservative but still physically-
motivated
priors; 
if more aggressive priors on nuisance parameters are used, the constraints are likely to improve further. 

In conclusion, we note that the addition of the one-loop bispectrum may yield some $\approx30\%$ improvement on the amplitude of equilateral PNG. We stress, however, that this comes with two important caveats. First, the $\kmax^B$ used for this study ($0.2\hMpc$) results in noticeable biases on the nuisance parameters, suggesting that the errorbar on $f_{\rm NL}^{\rm equil}$
may be underestimated due to over-fitting. Whether this induces a bias on $f_{\rm NL}^{\rm equil}$ is unclear; such an error would likely show up only in the analysis of simulations containing $f_{\rm NL}^{\rm equil}\neq 0$. Secondly, we have neglected the PNG-induced 
one-loop corrections to the bispectrum (as in \citep{DAmico:2022gki}), which can be marginally important for the scales of interest (particularly in the tails of the $f_{\rm NL}^{\rm equil}$ posterior), as can be easily estimated 
with the scaling universe approximation outlined in~\cite{Assassi:2015jqa}.

Finally, we note that our analysis indicates a more modest improvement on $f_{\rm NL}^{\rm equil}$ than that reported in
~\cite{DAmico:2022gki}. 
\cite{DAmico:2022gki} suggest that the one-loop
bispectrum improves 
$f_{\rm NL}^{\rm equil}$ constraints over the tree-level
result by a factor of few.
This follows from a comparison 
with the tree-level bispectrum analysis 
of~\cite{Cabass:2022wjy}. 
This comparison is misleading, however, 
since the baseline analysis of 
\cite{Cabass:2022wjy} varies cosmology whilst
\cite{DAmico:2022gki} always keeps cosmological
parameters fixed. We have checked that this accounts for most of the difference between 
\cite{Cabass:2022wjy} and \cite{DAmico:2022gki}.
A more detailed comparison with \cite{DAmico:2022gki}
is not currently possible because the former work has not yet presented sufficient details about their analysis and theory model. It will be interesting to compare our
results with their analysis in the future.

\section{Conclusions and Discussion}\label{sec: discussion}

In this work, we have presented and validated a complete calculation of the galaxy bispectrum monopole in redshift space at one-loop order in effective field theory. Our model includes one-loop corrections due to mode coupling, as well as the full set of EFT counterterms that are needed to regulate the UV behavior of loop integrals and capture the physical effects of backreaction of short scales onto the large-scale modes. Furthermore, we incorporate a bias expansion up to fourth order (noting that many operators vanish after renormalization of the power spectrum and bispectrum) as well as fourth order redshift-space distortions. In addition, our calculation includes IR resummation to capture the non-linear evolution of baryon acoustic oscillations (both for the power spectrum and bispectrum), as well as projection and binning effects. In short, we include all relevant ingredients needed to compare theory with observational galaxy clustering data. 

We have studied the performance of the one-loop bispectrum model in terms of cosmological parameter constraints, focusing primarily on the mass fluctuation amplitude $\sigma_8$. To validate our model we use the PT Challenge simulation suite \citep{Nishimichi:2020tvu}, which are equivalent to a BOSS-like survey with a hundred times larger volume, thus allowing for high-precision tests. We analyze a data vector that consists of the standard redshift-space power spectrum multipoles, the real-space power spectrum proxy $Q_0$ \citep{Ivanov:2021fbu}, and the redshift-space bispectrum monopole. In this setup, we have found that the inclusion of the one-loop corrections allows us to extend the agreement between bispectrum theory and data up to $\kmax =0.15\hMpc$, or $\kmax = 0.21\hMpc$ in real-space. This can be contrasted with the tree-level model bispectrum model, which works only up to $\kmax =0.08\hMpc$~\cite{Ivanov:2021kcd}. We caution that these scale-cuts depend on both the survey volume and galaxy type: for BOSS, we can use $\kmax = 0.20\hMpc$, and it is likely that the wavenumber reach is larger for emission line galaxies, which boast smaller FoG effects \cite{Chudaykin:2019ock,Ivanov:2021zmi}. Further, one might hope to extend the $k$-reach by specializing to some real-space bispectrum analog (similar to $Q_0$) at high-$k$: this will be considered in future work.

Despite a significant extension of the $k$-space reach, we have not found the bispectrum to lead to noticeable improvements in the $\sigma_8$ constraints compared to those with obtained from tree-level theory when applied to the PT Challenge simulations. This is a consequence of the large number of the EFT nuisance parameters that appear in the one-loop calculation (particularly in redshift-space), and must be marginalized over in our analysis. For a BOSS-volume survey (and accompanying systematic error thresholds), we find greater utility, with the one-loop bispectrum improving constraints by $\sim 10\%$ over the tree-level case, though it remains to be seen whether any accompanying shifts in nuisance parameters are real (and malignant) or just prior volume effects. 

We have additionally studied whether the one-loop bispectrum can help constrain equilateral primordial non-Gaussianity (and thus single-field inflation), finding that, for the BOSS survey, the one-loop bispectrum may improve constraints on the non-Gaussianity parameter $f_{\rm NL}^{\rm equil}$ by $\approx 30\%$ compared to the tree-level theory. Achieving this, however, requires pushing the bispectrum analysis to $\kmax^B=0.2\hMpc$, where the shifts in the bias parameters become evident. It remains to be seen if this problem can be alleviated with better priors on nuisance
parameters or with one-loop PNG-induced corrections to the bispectrum, which were omitted in this study. If one is interested in astrophysics, the one-loop bispectrum is much more useful: we find a significant tightening in the posteriors of parameters such as linear and tidal bias compared to those with only tree-level theory.

An important conclusion from our study is that we need better knowledge of the EFT nuisance parameters if we wish to extract more cosmological information from the bispectrum. This can be done in several ways. First, one can include data from the higher-order angular moments of the redshift-space bispectrum~\cite{Scoccimarro:1999ed}. Since these moments depend on the same set of parameters, their inclusion should tighten the EFT nuisance parameters posteriors, aiding determination of the cosmological parameters of interest. Second, one can constrain the EFT nuisance parameters with higher order statistics, such as the trispectrum, see e.g.~\cite{Bertolini:2016bmt,Philcox:2021hbm} for work in this direction. Finally, one can obtain better priors on the extra nuisance parameters using high fidelity N-body or hydrodynamical simulations. A powerful route by which to acheive this involves EFT field level techniques, 
see e.g.~\cite{Baldauf:2015zga,Lazeyras:2015lgp,Schmittfull:2018yuk,Abidi:2018eyd,Schmidt:2020viy,Schmittfull:2020trd,Steele:2020tak,Lazeyras:2021dar,Barreira:2021ukk,Steele:2021lnz}. 
We plan to investigate these options in future work. 

Though the one-loop bispectrum analysis of this work was limited only to two cosmological parameters, the mass fluctuation amplitude and the equilateral non--Gaussianity parameter, it may be similarly extended to other parameters such as local primordial non-Gaussianity, or the neutrino mass. The improvement on other parameters, especially those beyond the minimal $\Lambda$CDM model, could be significantly larger, particularly when some new feature is introduced that is not degenerate with the smooth loop corrections. If the parameter of interest enters the theoretical model linearly, the analysis can proceed as above; if this is not the case, one would require an optimization of our one-loop bispectrum pipeline, since the FFTLog-based approach does not currently allow for a fast re-calculation of the theoretical template as the power spectrum is varied. If only the $\alpha_{\parallel},\alpha_\perp$ parameters are varied however, the templates do not need to be recomputed, only rebinned (via \ref{eq: Bk-mon-AP}). Analyses including such effects will be natural next steps in our research program. 

Finally, we note that the bispectrum data offers novel probes of new physics. In particular, constructing the bispectrum from different tracers will allow one to probe the equivalence principle~\cite{Peloso:2013zw,Kehagias:2013yd,Valageas:2013cma,Creminelli:2013nua,Lewandowski:2019txi,Creminelli:2013mca,Creminelli:2013poa}. Such an analysis is complicated if one considers only the power spectrum since the effects sensitive to the equivalence principle appear there only at the one-loop order. In contrast, the cross-bispectrum of different kinds of tracers can be a sensible probe of the equivalence principle, whose violation would generate new bispectrum shapes that are not present in the $\Lambda$CDM model. This, in particular, will help one derive new constraints on the violation of Lorentz symmetry in the dark matter sector~\cite{Blas:2012vn,Audren:2014hza}. We leave this and other tests of new physics with the bispectrum for future work.

\paragraph{Acknowledgments} 
We would like to thank Emanuele Castorina, Azadeh Moradinezhad Dizgah and Zvonimir Vlah for useful discussions. OHEP thanks the Simons Foundation for support and acknowledges the Institute for Advanced Study for their hospitality and abundance of baked goods. The work of MMI has been supported by NASA through the NASA Hubble Fellowship grant \#HST-HF2-51483.001-A awarded by the Space Telescope Science Institute, which is operated by the Association of Universities for Research in Astronomy, Incorporated, under NASA contract NAS5-26555. GC acknowledges support from the Institute for Advanced Study. MZ is supported by the Canadian Institute for Advanced Research (CIFAR) program on Gravity and the Extreme
Universe and the Simons Foundation Modern Inflationary Cosmology initiative. This work was supported in part by MEXT/JSPS KAKENHI Grant Number JP19H00677, JP20H05861, JP21H01081 and JP22K03634. We also acknowledge financial support from Japan Science and Technology Agency (JST) AIP Acceleration Research Grant Number JP20317829.

The simulation data analysis was performed partly on Cray XC50 at Center for Computational Astrophysics, National Astronomical Observatory of Japan. Data analysis was primarily performed on the Helios cluster at the Institute for Advanced Study, Princeton, and additional computations were carried out using the Princeton Research Computing resources at Princeton University, which is a consortium of groups led by the Princeton Institute for Computational Science and Engineering (PICSciE) and the Office of Information Technology's Research Computing Division.

\appendix 

\section{Perturbation Theory Kernels}\label{app:kernels}

\subsection{Real-Space}\label{app:kernels-real}
At fourth-order, the real-space perturbation theory kernels (cf.\,\ref{eq: delta-n-exp}) are given by
\beq\label{eq: real-kernels}
    K_1(\vq_1) &=& \bb{b_1} , \\\nonumber
    K_2(\vq_1,\vq_2) &=&\left\{\bb{b_1}F_2(\vq_1,\vq_2)\right\}+\left\{\frac{\bb{b_2}}{2}+\bb{\gamma_2}\,\kappa(\vq_1,\vq_2)\right\} , \\\nonumber
    K_3(\vq_1,\vq_2,\vq_3)&=& \left\{\bb{b_1}F_3(\vq_1,\vq_2,\vq_3)\right\} + \left\{\bb{b_2}F_2(\vq_1,\vq_2)+2\bb{\gamma_2}\,\kappa(\vq_1,\vq_{23})G_2(\vq_2,\vq_3)\right\}\\\nonumber
    &&\,+\,\left\{\frac{\bb{b_3}}{6}+\bb{\gamma_2^\times}\,\kappa(\vq_1,\vq_2)+\bb{\gamma_3}\,L(\vq_1,\vq_2,\vq_3)+\bb{\gamma_{21}}\kappa(\vq_1,\vq_{23})\kappa(\vq_2,\vq_3)\right\} , \\\nonumber
    K_4(\vq_1,\vq_2,\vq_3,\vq_4)&=&\left\{\bb{b_1}F_4(\vq_1,\vq_2,\vq_3,\vq_4)\right\}\\\nonumber
    &&\,+\,\left\{\frac{\bb{b_2}}{2}\left[F_2(\vq_1,\vq_2)F_2(\vq_3,\vq_4)+2F_3(\vq_1,\vq_2,\vq_3)\right]\right.\\\nonumber
    &&\qquad\left.+\,\bb{\gamma_2}\left[\kappa(\vq_{12},\vq_{34})G_2(\vq_1,\vq_2)G_2(\vq_3,\vq_4)+2\kappa(\vq_{123},\vq_4)G_3(\vq_1,\vq_2,\vq_3)\right]\right\}\\\nonumber
    &&\,+\,\left\{\frac{\bb{b_3}}{2}F_2(\vq_1,\vq_2)+\bb{\gamma_2^\times}\left[2\kappa(\vq_{12},\vq_3)G_2(\vq_1,\vq_2)+\kappa(\vq_3,\vq_4)F_2(\vq_1,\vq_2)\right]\right.\\\nonumber
    &&\qquad+\,3\bb{\gamma_3}\,L(\vq_1,\vq_2,\vq_{34})G_2(\vq_3,\vq_4)\\\nonumber
    &&\qquad\left.+\,\bb{\gamma_{21}}\left[\kappa(\vq_{12},\vq_{34})\kappa(\vq_1,\vq_2)F_2(\vq_3,\vq_4)+2\kappa(\vq_{123},\vq_4)\kappa(\vq_{12},\vq_3)F_2(\vq_1,\vq_2)\right]\right\}\\\nonumber
    &&\,+\,\left\{\bb{\gamma_{21}^\times}\,\kappa(\vq_1,\vq_{23})\kappa(\vq_2,\vq_3)+\bb{\gamma_{211}}L(\vq_1,\vq_2,\vq_{34})\kappa(\vq_3,\vq_4)\right.\\\nonumber
    &&\qquad+\,\bb{\gamma_{22}}\,\kappa(\vq_{12},\vq_{34})\kappa(\vq_1,\vq_2)\kappa(\vq_3,\vq_4)\\\nonumber
    &&\qquad+\,\bb{\gamma_{31}}\left[\frac{1}{18}\kappa(\vq_1,\vq_{234})\left(\frac{15}{7}\kappa(\vq_{23},\vq_4)\kappa(\vq_2,\vq_3)-L(\vq_2,\vq_3,\vq_4)\right)\right.\\\nonumber
    &&\qquad\qquad\left.\left.+\frac{1}{14}\left(M(\vq_1,\vq_{23},\vq_4,\vq_{234})-M(\vq_1,\vq_{234},\vq_{23},\vq_4)\right)\kappa(\vq_2,\vq_3)\right]\right\},
\eeq
where the bias parameters are shown in color, $\vq_{i\cdots j}\equiv \vq_i+\cdots+\vq_j$ and we define the angles
\beq\label{eq: ang-def}
    \kappa(\vq_1,\vq_2) &=& (\hq_1\cdot\hq_2)^2-1 , \\\nonumber
    L(\vq_1,\vq_2,\vq_3) &=& 2(\hq_1\cdot\hq_2)(\hq_2\cdot\hq_3)(\hq_3\cdot\hq_1)-(\hq_1\cdot\hq_2)^2-(\hq_2\cdot\hq_3)^2-(\hq_3\cdot\hq_1)^2+1 , \\\nonumber
    M(\vq_1,\vq_2,\vq_3,\vq_4) &=& (\hq_1\cdot\hq_2)(\hq_2\cdot\hq_3)(\hq_3\cdot\hq_4)(\hq_4\cdot\hq_1).
\eeq
\eqref{eq: real-kernels} uses the standard Eulerian density and velocity kernels ($F_n$ and $G_n$) and we note that the kernels have not been symmetrized over their arguments. We additionally note that, despite appearing in $K_4$, $b_3$ does not contribute to the one-loop bispectrum after averaging over internal angles and performing bias renormalization. Furthermore, we drop any bias operators in $K_4$ that do not appear in the one-loop bispectrum, thus the above expression will not be sufficient for computation of the two-loop power spectrum.

\subsection{Redshift-Space}\label{app:kernels-rsd}
The redshift-space kernels are obtained by expanding the RSD mapping of \eqref{eq: rsd-mapping} and expanding all fields in terms of the linear density field. Following a lengthy computation, we find the following forms in terms of the real-space kernels:
\beq\label{eq: RSD-kernels}
    Z_1(\vq_1) &=& K_1 + f\mu_1^2,\\\nonumber
    Z_2(\vq_1,\vq_2) &=& K_2(\vq_1,\vq_2) + f\mu_{12}^2G_2(\vq_1,\vq_2)+\frac{f\mu_{12}q_{12}}{2}K_1\left[\frac{\mu_1}{q_1}+\frac{\mu_2}{q_2}\right]+\frac{(f\mu_{12}q_{12})^2}{2}\frac{\mu_1}{q_1}\frac{\mu_2}{q_2} , \\\nonumber
    Z_3(\vq_1,\vq_2,\vq_3) &=& K_3(\vq_1,\vq_2,\vq_3)+f\mu_{123}^2G_3(\vq_1,\vq_2,\vq_3)\\\nonumber
    &&\,+\,(f\mu_{123}q_{123})\left[\frac{\mu_{12}}{q_{12}}K_1G_2(\vq_1,\vq_2)+\frac{\mu_3}{q_3}K_2(\vq_1,\vq_2)\right]\\\nonumber
    &&\,+\,\frac{(f\mu_{123}q_{123})^2}{2}\left[2\frac{\mu_{12}}{q_{12}}\frac{\mu_3}{q_3}G_2(\vq_1,\vq_2)+\frac{\mu_1}{q_1}\frac{\mu_2}{q_2}K_1\right]+\frac{(f\mu_{123}q_{123})^3}{6}\frac{\mu_1}{q_1}\frac{\mu_2}{q_2}\frac{\mu_3}{q_3} , \\\nonumber
    Z_4(\vq_1,\vq_2,\vq_3,\vq_4) &=&  K_4(\vq_1,\vq_2,\vq_3,\vq_4)+f\mu_{1234}^2G_4(\vq_1,\vq_2,\vq_3,\vq_4)\\\nonumber
    &&\,+\,(f\mu_{1234}q_{1234})\left[\frac{\mu_{123}}{q_{123}}K_1G_3(\vq_1,\vq_2,\vq_3)+\frac{\mu_4}{q_4}K_3(\vq_1,\vq_2,\vq_3)\right.\\\nonumber
    &&\qquad\qquad\qquad\qquad+\,\left.\frac{\mu_{12}}{q_{12}}G_2(\vq_1,\vq_2)K_2(\vq_3,\vq_4)\right]\\\nonumber
    &&\,+\,\frac{(f\mu_{1234}q_{1234})^2}{2}\left[2\frac{\mu_{123}}{q_{123}}\frac{\mu_4}{q_4}G_3(\vq_1,\vq_2,\vq_3)+\frac{\mu_{12}}{q_{12}}\frac{\mu_{34}}{q_{34}}G_2(\vq_1,\vq_2)G_2(\vq_3,\vq_4)\right.\\\nonumber
    &&\qquad\qquad\qquad\qquad\left.+\,2\frac{\mu_{12}}{q_{12}}\frac{\mu_3}{q_3}K_1G_2(\vq_1,\vq_2)+\frac{\mu_1}{q_1}\frac{\mu_2}{q_2}K_2(\vq_3,\vq_4)\right]\\\nonumber
    &&\,+\,\frac{(f\mu_{1234}q_{1234})^3}{6}\left[3\frac{\mu_{12}}{q_{12}}\frac{\mu_3}{q_3}\frac{\mu_4}{q_4}G_2(\vq_1,\vq_2)+\frac{\mu_1}{q_1}\frac{\mu_2}{q_2}\frac{\mu_3}{q_3}K_1\right]\\\nonumber
    &&\,+\,\frac{(f\mu_{1234}q_{1234})^4}{24}\frac{\mu_1}{q_1}\frac{\mu_2}{q_2}\frac{\mu_3}{q_3}\frac{\mu_4}{q_4},
\eeq
writing $\mu_{i\cdots j}\equiv \mu_{\vq_i+\cdots+\vq_j}$ and dropping the argument of $K_1$ for clarity.

\section{Computation of the One-Loop Bispectrum with FFTLog}\label{app:fftlog}
In this appendix, we discuss practical computation of the loop integrals given in \eqref{eq: Bk-one-loop}. Before considering the redshift-space case, we will first examine how to compute the real-space integrals, which follow a similar logic, but are significantly simpler. Our approach follows \citep{Simonovic:2017mhp}, but is extended to the case of biased tracers and redshift-space.

\subsection{Real-Space}\label{app:fftlog-real}
\subsubsection{Formalism} 
As noted in \S\ref{sec: theory}, the first step in the bispectrum computation is the expansion of the perturbation theory kernels (Appendix \ref{app:kernels}) as polynomials in $q^2$, $|\vk_1-\vq|^2$ and $|\vk_2+\vq|^2$ or their reciprocals (utilizing permutation symmetries). In practice, this results in a sum over many thousands of terms, once the relevant symmetries have been imposed, and is automated using \textsc{mathematica}. For $B_{222}$, each term is proportional to
\beq\label{eq: fft-log-demo}
    \int_{\vq}\frac{P_L(q)P_L(|\vk_1-\vq|)P_L(|\vk_2+\vq|)}{q^{\alpha_1}|\vk_1-\vq|^{\alpha_2}|\vk_2+\vq|^{\alpha_3}},
\eeq
for integer $\alpha_i$, with a similar form found for the other loop integrals except with fewer factors of $P_L$. Expanding the linear power spectrum as a sum of complex polynomials in $k$, \textit{i.e.}\ $P_L(k)= \sum_m c_m\,k^{\nu+i\eta_m}$ for frequencies $\eta_m$, coefficients $c_m$, and (real) FFTLog bias $\nu$ (which sets the eventual integral convergence properties), we can rewrite \eqref{eq: fft-log-demo} in the form
\beq\label{eq: fft-log-master-app}
    \sum_{m_1m_2m_3}c_{m_1}c_{m_2}c_{m_3}\int_{\vq}\frac{1}{q^{2\nu_1}|\vk_1-\vq|^{2\nu_2}|\vk_2+\vq|^{2\nu_3}}
\eeq
for $2\nu_j=\alpha_j-\nu-j\eta_{m_i}$, where all the cosmology dependence (encoded in $c_m$) is now outside the integral. The remaining integral can be computed using path integral methods as
\beq\label{eq: JC-def}
    \int_{\vq}\frac{1}{q^{2\nu_1}|\vk_1-\vq|^{2\nu_2}|\vk_2+\vq|^{2\nu_3}} \equiv k^{3-2(\nu_1+\nu_2+\nu_3)}\mathsf{J}(\nu_1,\nu_2,\nu_3;x,y)
\eeq
where $x^2=k_3/k_1$, $y^2=k_2/k_1$ and $\mathsf{J}$ (with complex arguments $\nu_i$) can be expressed as a sum of hypergeometric functions and Gamma functions \citep{Simonovic:2017mhp}. This reduces the computation of bispectrum templates to a set of matrix multiplications and function evaluations, as noted in \S\ref{sec: implementation}. For $B_{321}^{I}$ we find a similar form to \eqref{eq: fft-log-master}, except with rank-two matrices, whilst $B_{321}^{II}$ and $B_{411}$ involve only a one-dimensional sum (and one set of $c_{m_i}$ coefficients).

\subsubsection{Limiting Behavior}
When computing spectra via FFTLog, it is important to verify whether the relevant loop integrals actually converge. This is achieved by taking the UV and IR limits of the integration kernels and assessing the dependence on the $\nu_i$ parameters appearing in \eqref{eq: fft-log-master-app}. As an example, we consider the contribution of three $\delta$ operators to $B_{222}$ (involving three copies of $b_1F_2(\vk-\vq,\vq)$). This has the following limits in the equilateral configuration $k_1\sim k_2\sim k_3\sim k$:
\beq
    B_{222}^{\rm UV}(\vk_1,\vk_2,\vk_3)&\supset& \frac{89}{2401}\bb{b_1}^3\,k^6\int_{\vq}\frac{P_L^3(q)}{q^6},\\\nonumber
    B_{222}^{\rm IR}(\vk_1,\vk_2,\vk_3)&\supset& \frac{2}{21}\bb{b_1}^3k^2P_L^2(k)\int_{\vq}\frac{P_L(q)}{q^2}.
\eeq
For $P_L(q)\sim q^\nu$, the integral is UV convergent for $\nu<1$ and IR convergent for $\nu>-1$. By choosing the bias in this range, FFTLog will give accurate values for the integrals. In contrast, if $\nu$ is chosen to be outside this range, we must add the relevant UV or IR limits by hand (taking care to include subleading divergences if necessary).

Considering all bias terms, the limits of $B_{222}$ take the following schematic form for equilateral triangles:
\beq\label{eq: B222-UV-IR}
    B_{222}^{\rm UV} &\sim& \int_\vq P_L^3(q)\\\nonumber
    &&\,\times\,\left\{\bb{b_2}^3+\left(\frac{k}{q}\right)^2\bb{b_2}^2f_1(\bb{b_1},\bb{\gamma_2})+\left(\frac{k}{q}\right)^4\bb{b_2}\,f_2(\bb{b_1},\bb{\gamma_2})+\left(\frac{k}{q}\right)^6f_2(\bb{b_1},\bb{\gamma_2})\right\}\\\nonumber
    B_{222}^{\rm IR} &\sim& P^2_L(k)\int_\vq P_L(q)\left\{\left(\frac{k}{q}\right)^2\left[\bb{b_1}^3f_4+\bb{b_1}^2f_5(\bb{b_2},\bb{\gamma_2})\right]+\left[\bb{b_1}f_6(\bb{b_2},\bb{\gamma_2})+f_7(\bb{b_2},\bb{\gamma_2})\right]\right\},
\eeq
where $\{f_i\}$ are some polynomials, and we consider only the leading-order contribution for each bias parameter. Inserting $P_L(q)\sim q^\nu$ as before shows that a term containing $K$ powers of $b_2$ is UV convergent for $\nu<1-2K/3$, implying $\nu<-1$ for $b_2^3$, significantly tighter than the $\nu<1$ limit for matter (\textit{i.e.}\ $b_1^3$). However, the UV limit of $b_2^3$ is fully degenerate with the bispectrum shot-noise ($\epsilon_0$ in \ref{eq: Bstoch}), and should be subtracted off in practice, as for the $b_2^2$ contribution to $P_{22}$. If we adopt $\nu>-1$, this term will not be captured by the FFTLog formalism, thus the subtraction becomes implicit. In this case, we require $\nu<-1/3$ to avoid the $b_2^2$ divergence (and the second-order $b_2^3$ divergence, both of which are degenerate with the $\epsilon_2$ stochasticity in \ref{eq: Bstoch}). In the IR, \eqref{eq: B222-UV-IR} shows that the integral is convergent for $\nu>-1$ for terms involving two or more powers of $b_1$, and $\nu>-3$ else. To satisfy all the conditions simultaneously, we may take $-1<\nu<-1/3$, dropping the shot-noise piece. 

For $B_{321}^I$, the limiting UV and IR form is given by
\beq\label{eq: B321I-UV-IR}
    B_{321}^{I,\rm UV} &\sim& \bb{b_1}P_L(k)\int_\vq P^2_L(q)\left\{\frac{1}{2}\bb{X}^2+\left(\frac{k}{q}\right)^2\bb{X}f_1(\cdots)+\left(\frac{k}{q}\right)^4f_2(\cdots)\right\}\\\nonumber
    B_{321}^{I,\rm IR} &\sim& \bb{b_1}P_L^2(k)\int_\vq P_L(q)\left\{\left(\frac{k}{q}\right)^2\left[\bb{b_1}^2f_3(\cdots)+\bb{b_1}f_4(\cdots)\right]+f_6(\cdots)\right\},
\eeq
where $X\in \{b_2,b_3,\gamma_2^\times\}$ and ellipses are taken to mean bias operators excluding $X$ (in the UV) or $b_1$ (in the IR). \eqref{eq: B321I-UV-IR} implies that UV divergences can be avoided if we take $\nu<(1-2K)/2$ when the term involves $K$ powers of $b_2$, $b_3$, or $\gamma_2^\times$; these are all the composite operators appearing at third order. Furthermore, as in $B_{222}$, the UV limits of the terms involving two powers of $b_2$, $b_3$ and $\gamma_2^\times$ are proportional to shot-noise (this time of the $\eta_0$ variety in \ref{eq: Bstoch}),\footnote{This occurs regardless of the triangle configuration, once permutations are taken into account.} and should be subtracted off in practice (or dropped implicitly by fixing $\nu>-3/2$). In the IR, divergences vanish for $\nu>-1$ for terms involving $b_1^3$ or $b_1^2$ and $\nu>-3$ else. Overall, we require a bias of $-1<\nu<-1/2$ to satisfy all conditions, assuming subtraction of the $\eta_0$ shot-noise contributions. 

For $B_{321}^{II}$, we require the UV and IR convergence properties of $\int_{\vq}Z_3(\vk,\vq,-\vq)P_L(q)$, which we label $\tilde P_{13}(k)$ by analogy with the galaxy power spectrum \eqref{eq: Bk-one-loop}. This natively involves all bias operators in \eqref{eq: bias-expan-4th} up to third order; however, this set is reduced to just $\{\delta,\cG_2(\Phi_v),\cG_2(\varphi_2,\varphi_1)\}$ when the renormalization conditions are applied. These conditions demand
\beq
    \lim_{k_{1}\to0}\av{X(\vk)\delta_L(\vk_1)}'_{\rm loop} = 0
\eeq
for renormalized operator $X$ and linear density field $\delta_L(\vk)$, \textit{i.e.}\ there can be no loop contributions which do not decay in the UV limit \citep{Assassi:2014fva}. The contribution of all composite operators (e.g., $\delta^2$) to $\tilde P_{13}$ is exactly that of a non-decaying loop diagram (since there is no suppression by the $F_3$ kernel), thus must vanish when the operators are properly renormalized. This leaves only $\mathcal{G}_3$, which evaluates to zero after averaging over the angular part of $\vq$. Following these redefinitions (which do not affect $B_{222}$ and $B_{321}^I$), we find the UV and IR limits:
\beq\label{eq: P13-UV-IR}
    \tilde P_{13}^{\rm UV} &\sim& \int_{\vq}P_L(q)\left(\frac{k}{q}\right)^2f_1(\bb{b_1},\bb{\gamma_2},\bb{\gamma_{21}})\\\nonumber
    \tilde P_{13}^{\rm IR} &\sim& \int_{\vq}P_L(q)\left\{\left(\frac{k}{q}\right)^2\bb{b_1}f_2+f_3(\bb{\gamma_2},\bb{\gamma_{21}})\right\}.
\eeq
UV divergences occur unless $\nu<-1$, and IR divergences occur unless $\nu>-1$ (for $b_1$) or $\nu>-3$ (else). As for $P_{13}$ \citep{Simonovic:2017mhp}, there is no range of biases which satisfy all the conditions; in this case, we can choose $-1<\nu<1$ (satisfying the IR limits, and avoiding subleading UV divergences at $\nu>1$), and correct the UV part by adding the relevant limit by hand, which takes the following explicit form in real-space:
\beq
    \tilde P_{13}^{\rm UV}(k) = -k^2\left(\frac{61}{315}\bb{b_1}+\frac{64}{35}\bb{\gamma_2}-\frac{32}{15}\bb{\gamma_{21}}\right)\int_{0}^\infty \frac{q^2dq}{2\pi^2}\,\frac{P(q)}{q^2},
\eeq
proportional to the velocity divergence $\sigma^2_v$.

Finally, we consider $B_{411}$. This contains the fourth-order bias operators, and involves Wick contractions of linear density fields within the same operator, permitting simplification via the renormalization condition:
\beq\label{eq: renorm-B411}
    \lim_{k_{1,2}\to0}\av{X(\vk)\delta_L(\vk_1)\delta_L(\vk_2)}'_{\rm loop} = 0,
\eeq
which is proportional to the UV limit of $B_{411}$. The first effect of this is to remove contributions from any fourth-order composite local evolution operator (such as $\delta^4$ or $\delta\,\cG_3$); these operators were already dropped from the bias expansion in \eqref{eq: bias-expan-4th}. Secondly, this will remove a number of UV divergences in the below. Before bias renormalization, the UV and IR limits of the remaining terms take the form:
\beq\label{eq: B411-UV-IR}
    B_{411}^{\rm UV} &\sim& \bb{b_1}^2P_L^2(k)\int_\vq P_L(q)\left\{f_1(\bb{b_2},\bb{\gamma_2^\times},\bb{\gamma_{21}^\times})+\left(\frac{k}{q}\right)^2f_2(\cdots)\right\}\\\nonumber
    B_{411}^{\rm IR} &\sim& \bb{b_1}^2P_L^2(k)\int_\vq P_L(q)\left\{\left(\frac{k}{q}\right)^2f_3(\bb{b_1},\bb{b_2},\bb{\gamma_2})+f_4(\cdots)\right\},
\eeq
where ellipses represent additional bias terms which impatience lead us to ignore. The first line is UV convergent for $\nu<-3$ (first term, involving composite operators) or $\nu<-1$ (second term, no composite operators). However, the first term possesses a UV limit that does not decay as an (negative) integer power of $q^2$, violating the renormalization condition \eqref{eq: renorm-B411}. The precise action of bias operator renormalization is to remove such terms (and only these, as far as this diagram is concerned). By evaluating the diagram with $\nu>-1$, such contributions will be avoided, \textit{i.e.}\ the operators will be correctly renormalized. In the IR, we find that divergences can be avoided by setting $\nu>-1$ (for terms involving the first and second order operators proportional to $b_1$, $\gamma_2$ or $b_2$), or $\nu>-3$ (for the remaining terms). As for $\tilde P_{13}$, there is no single bias that will simultaneously remove all the UV and IR divergences in $B_{411}$, even after bias renormalization. Fixing $-1<\nu<1$, we may compute the full expression by manually adding the appropriate UV limit to the FFTLog result. These limits can be computed straightforwardly from the kernels in \textsc{mathematica} and are omitted from this publication to avoid unnecessary tedium.

\subsection{Redshift-Space}\label{subsec: computation-rsd}
In redshift-space the perturbation theory kernels depend not only on the lengths $q$, $|\vk_1-\vq|$ and $|\vk_2+\vq|$ but also the LoS angles $\mu_i\equiv \hk_i\cdot\hn$ and $\hq\cdot\hn$.\footnote{In the isotropic approximation of infrared resummation (\S\ref{sec: theory}) the angular dependence is purely polynomial.} Although we are primarily interested only in the bispectrum monopole (\textit{i.e.}\ that integrated over $\mu_{1,2}$, with a suitable Lebesque measure, as in \ref{eq: Bk-mon}), the full dependence on $\mu_i$ is necessary for accurate calculation of coordinate distortions \eqref{eq: Bk-mon-AP}, thus we cannot simply average over $\mu_i$ before computing the loop integrals; furthermore, this is difficult to perform analytically due to the presence of high powers of $\hq\cdot\hn$. After expanding the kernels as polynomials, we will find loop integrals of the form:
\beq\label{eq: fftlog-rsd}
    \int_{\vq}\frac{(\hq\cdot\hn)^n}{q^{2\nu_1}|\vk_1-\vq|^{2\nu_2}|\vk_2+\vq|^{2\nu_3}}
\eeq
for $n\in\{0,1,\cdots,6\}$ (cf.\,\ref{eq: fft-log-master}), with prefactors depending on $\mu_i$, $k_i$, biases and $f(z)$. Below, we consider how to compute this utilizing the FFTLog procedure, generalizing the approach of \citep{Chudaykin:2020aoj} for the power spectrum.

First, we expand the $\hq\cdot\hn$ angles as Cartesian sums, \textit{i.e.}\ $\sum_{i=1}^3 \hat{q}_i\hat{n}^i$, and pull out the LoS vectors from the integral. The remaining function is a fully symmetric rank-$n$ tensor, given by
\beq\label{eq: f-matrix}
    \mathcal{F}^{i_1\cdots i_n}\equiv \int_{\vq}\frac{\hat q^{i_1}\cdots\hat q^{i_n}}{q^{2\nu_1}|\vk_1-\vq|^{2\nu_2}|\vk_2+\vq|^{2\nu_3}}.
\eeq
This has dependence only on $\vk_1$ and $\vk_2$; as such, its tensorial dependence can be written in terms of the components of $\vk_1$, $\vk_2$, and any isotropic tensors of relevance, \textit{i.e.}\ the Kronecker delta.\footnote{Note that the Levi-Cevita tensor, which is relevant for rank-three tensors and above, does not contribute to the expansion of \eqref{eq: f-matrix}, since it is antisymmetric.} Explicitly, this takes the form:
\beq\label{eq: fftlog-rsd-general-n}
     \mathcal{F}^{i_1\cdots i_n}=\sum_k A_k\,\mathcal{O}_k^{i_1\cdots i_n},
\eeq
where $\{\mathcal{O}_k\}$ is the set of all independent symmetric rank-$n$ combinations of $\hat k_1^i$, $\hat k_2^i$ and $\delta_{\rm K}^{ij}$. As an example, the $n=2$ operators are $\{\delta_{\rm K}^{ij},\hat k_1^i\hat k_1^i, \hat k_1^i\hat k_2^j+\hat k_2^i\hat k_1^j, \hat k_2^i\hat k_2^j\}$. We then define an ``overlap matrix'', giving the correlation between basis elements:
\beq
    \mathcal{I}_{kk'} = \mathcal{O}_k^{i_1\cdots i_n}\mathcal{O}^{k'}_{i_1\cdots i_n}
\eeq
(assuming Einstein summation conventions); this allows extraction of the $A_k$ coefficients via $A_k = \left[\mathcal{I}\right]^{-1}_{kk'}\mathcal{O}^{k'}_{i_1\cdots i_n}\mathcal{F}^{i_1\cdots i_n}$, where the second term is just the contraction of \eqref{eq: fftlog-rsd-general-n} with various powers of $\hk_1$ and $\hk_2$.
Finally, we contract \eqref{eq: fftlog-rsd-general-n} with $n$ copies of $\hat n_i$ to yield
\beq\label{eq: rsd-trick}
    \int_{\vq}\frac{(\hq\cdot\hn)^n}{q^{2\nu_1}|\vk_1-\vq|^{2\nu_2}|\vk_2+\vq|^{2\nu_3}} = \left(\left[\mathcal{I}\right]^{-1}_{kk'}\mathcal{O}^{k'}_{i_1\cdots i_n}\mathcal{F}^{i_1\cdots i_n}\right)\left(\mathcal{O}_{k}^{i_1\cdots i_n}\hat{n}_{i_1}\cdots\hat{n}_{i_n}\right),
\eeq
where the first set of parentheses contains a set of $(\hk_{1,2}\cdot\hq$ coefficients inside the $\vq$ integral, and the second contains powers of $\mu_{1,2}$. To make this explicit, we give the $n=1$ case:
\beq\label{eq: n=1-fftlog}
    \int_{\vq}\frac{(\hq\cdot\hn)}{q^{2\nu_1}|\vk_1-\vq|^{2\nu_2}|\vk_2+\vq|^{2\nu_3}} &=& \frac{\mu_1-\mu_2\nu_{12}}{1-\nu_{12}^2}\int_{\vq}\frac{\hq\cdot\hk_1}{q^{2\nu_1}|\vk_1-\vq|^{2\nu_2}|\vk_2+\vq|^{2\nu_3}}\\\nonumber
    &&\,+\,\frac{\mu_2-\mu_1\nu_{12}}{1-\nu_{12}^2}\int_{\vq}\frac{\hq\cdot\hk_2}{q^{2\nu_1}|\vk_1-\vq|^{2\nu_2}|\vk_2+\vq|^{2\nu_3}},
\eeq
writing $\nu_{12}\equiv \hk_1\cdot\hk_2$. In this manner, the FFTLog integral can be performed for arbitrarily large $n$. We adopt this method to compute the bispectrum templates in redshift space, applying it as a simplification step before the loop integrals are computed as in Appendix \ref{app:fftlog-real}.

Notably, the above decomposition breaks down in the limit of $\hk_1\cdot\hk_2\to-1$, \textit{i.e.}\ for $\vk_1\parallel\vk_2$, whence there is only one angle in the problem. The corresponds to flattened triangles (with $k_1=k_2+k_3$ or $\sqrt{x}+\sqrt{y}=1$), which contain the divergence $1-\nu_{12}^2\to 0$. Strictly speaking, this divergence is cancelled by the numerators, once $\hk_1=-\hk_2$ is identified; however, if one separately computes the loop integral coefficients proportional to powers of $\mu_1$ and $\mu_2$, numerical issues will arise. In this limit, we adopt a different angular decomposition, noting that \eqref{eq: f-matrix} can depend only on $\hk_1$ and the Kronecker delta. The basis tensors are much simpler in this case, for example, with $\{\delta_{\rm K}^{ij},\hat k_1^i\hat k_1^j\}$ for $n=2$, and facilitate computation in an analogous manner to the above. For $n=1$, \eqref{eq: n=1-fftlog} becomes
\beq
    \left.\int_{\vq}\frac{(\hq\cdot\hn)}{q^{2\nu_1}|\vk_1-\vq|^{2\nu_2}|\vk_2+\vq|^{2\nu_3}}\right|_{\rm flat} &=& \mu_1\int_{\vq}\frac{\hq\cdot\hk_1}{q^{2\nu_1}|\vk_1-\vq|^{2\nu_2}|\vk_2+\vq|^{2\nu_3}},
\eeq
which does not diverge. This divergence also illustrates the importance of expanding the bispectrum templates in the $\{\mu,\chi\}$ basis (with $\mu\equiv \mu_1,\chi\equiv\sqrt{1-\mu^2}\cos\phi$) rather than $\{\mu_1,\mu_2\}$ (cf.\,\S\ref{sec: implementation}): the former is undefined for flattened triangles, whence $\mu_1=-\mu_2$, whilst the latter simply has dependence only on $\mu$ in this limit (noting that $\mu_2=\mu\,\nu_{12}-\sqrt{1-\nu_{12}^2}\,\chi \to -\mu$ as $\nu_{12}\to-1$).

\subsection{Implementation}\label{app:fftlog-imp}
The above tricks allow us to efficiently compute the one-loop bispectrum in redshift space. A rough overview of the computation is the following:
\begin{enumerate}
    \item Expand the relevant (symmetrized) perturbation theory kernels as polynomials in $q$, $|\vk_i\pm\vq|$, $\mu_i$ and $\hq\cdot\hn$.
    \item Replace powers of $\hq\cdot\hn$ with \eqref{eq: rsd-trick} and expand.
    \item Create lookup tables giving each $\{\nu_1,\nu_2,\nu_3\}$ triplet of FFTLog frequencies and the associated coefficient (depending on biases and $\mu_i$).
    \item Switch from the native $\{\mu_1,\mu_2,\mu_3\}$ basis to the (divergence-free) $\{\mu,\chi\}$ angular basis.
    \item Compute the bispectrum templates for each of the 47 combinations of $\mu^i\chi^j$ and the relevant combinations of bias parameters using the FFTLog algorithm. This is performed for a grid of values of $k_1, x, y$, with flattened templates (obeying $\sqrt{x}+\sqrt{y}=1$) computed separately, using the alternate angular decomposition given in Appendix \ref{subsec: computation-rsd}, and involving only 7 non-trivial powers of $\mu$.
    \item Create a three-dimensional linear interpolator for each template using the precomputed bispectrum shapes (combining full and flattened configurations).
    \item Apply binning and angular integration using \eqref{eq: bk-binning}, including the Alcock-Paczynski parameters if necessary.
    \item Compute the full bispectrum as a sum over templates, weighted by the bias configurations and any necessary discreteness weights.
\end{enumerate}
Notably, only steps (5) and beyond depend on the power spectrum template, and thus the cosmological survey in question. We note one further subtlety: computing the bispectrum templates near (but not at) the flattened limit of $\sqrt{x}+\sqrt{y}=1$ can lead to numerical issues due to large values of $1/(1-\nu_{12}^2)$, which appear in the angular decompositions of $(\hq\cdot\hn)$ raised to the $n$-th power (cf.\,\ref{eq: n=1-fftlog}). To counter this, when the templates are being computed and the condition $\sqrt{x}+\sqrt{y}<1.1$ is met, we replace the FFTLog prefactor by its Taylor series in $(1+\nu_{12})$, artificially removing the divergent terms (which are present only due to numerical inaccuracies). \textsc{Mathematica} and \textsc{Python} code implementing all of the above steps is publicly available at \href{https://github.com/oliverphilcox/OneLoopBispectrum}{GitHub.com/OliverPhilcox/OneLoopBispectrum}.

Following initial testing of the FFTLog routines against explicit numerical integration for a small number of bins, we use the following choices of FFTLog bias: $\nu = -0.6$ for $B_{222}$ and $B_{321}^{II}$, $\nu = -0.3$ for the matter terms in $B_{321}^{I}$ (\textit{i.e.}\ those involving only $b_1$ and $f$) and $B_{411}$, and $\nu = -0.8$ for the remaining terms in $B_{321}^{I}$. We additionally use $64$ FFTLog frequencies for $B_{222}$, $96$ frequencies for $B_{321}^{I}$, and $128$ frequencies for $B_{321}^{II}$ and $B_{411}$, which reduces computational costs, whilst keeping good resolution on the terms with enhanced oscillatory behavior. Additionally, we manually add the UV limits of $B_{321}^{II}$ and $B_{411}$ (as in Appendix \ref{app:fftlog-real}), and choose the $k$-range for the FFTLog decomposition to lie in $[10^{-5},10]\hMpc$, computing all shapes on an interpolation grid of $64$ $k_1$-values logarithmically spaced in $[10^{-3},0.5]\hMpc$, and $41$ equally-spaced values of each of $\sqrt{x}$ and $\sqrt{y}$, subject to the triangle conditions. This is sufficient to ensure that the spectra are subpercent accurate in the regime of interest; the results are largely unchanged if the number of FFTLog frequencies is reduced by a factor of two. The computation requires $\sim 10^4$ CPU-hours to compute all templates (entirely performed within \textsc{Mathematica}), with the majority of time devoted to $B_{222}$, and could certainly be optimized further. Calculations have been compared against explicit numerical integration of the (unsimplified) bispectrum kernels, and we find excellent sub-percent agreement in all cases.

\section{Counterterms from Redshift-Space Distortions}
\label{app:ctrs}

The RSD mapping to $\mathcal{O}(\delta_1^4)$ can be obtained by expanding \eqref{eq: rsd-mapping} to fourth order:
\be
\begin{split}
\delta_\k^{(s)} = & \delta_\k -ik_z[(1+\delta_g)u_z]_\k
+\frac{i^2}{2}k_z^2 [(1+\delta_g)u^2_z]_\k \\
& -\frac{i^3}{3!} k_z^3[(1+\delta_g)u^3_z]_\k 
+ \frac{i^4}{4!} k_z^4 [v_z^4(1+\delta_g)]_\k \,,
 \end{split} 
\ee
where $A_z\equiv \vec{A}\cdot \hat{n} \equiv A\mu_{\vec A}$. To facilitate renormalization, we must smooth this expansion
with a low-pass filter of some size $R=\Lambda^{-1}$. Products of fields at the same point (contact terms) are sensitive to short-scaled modes and hence must to be smoothed and renormalized.  We denote these operations by square brackets, $[...]_R$.
Galilean symmetry implies the following schematic structure of the renormalized correlators (see~\cite{Senatore:2014vja,Lewandowski:2015ziq,Perko:2016puo} for the first order results),
\be 
\begin{split}
& [u^i(1+\delta_g)]_R = (1+\delta_{g}{}_\ell)u^i_\ell +\mathcal{O}^i_u\,, \\
& [u^i v^j(1+\delta_g)]_R = (1+\delta_{g}{}_\ell)u^i_\ell u^j_\ell +u^i_\ell \mathcal{O}^{(1)}{}^j_{u^2} 
+ u^j_\ell \mathcal{O}^{(1)}{}^i_{u^2} + \mathcal{O}^{(2)}{}^{ij}_{u^2} \,,\\
& [u^i u^j u^k(1+\delta_g) ]_R = (1+\delta_{g}{}_\ell)
u^i_\ell 
u^j_\ell
u^k_\ell  
+3 u^2_\ell \mathcal{O}_{u^3}^{(1)ijk} 
+ 3 u_\ell \mathcal{O}_{u^3}^{(2)ijk} 
+ \mathcal{O}_{u^3}^{(3)ijk} \,,\\
& [u^i u^j u^k u^l (1+\delta_g)]_R = 
(1+\delta_{g}{}_\ell)
u^i_\ell 
u^j_\ell 
u^k_\ell 
u^l_\ell 
+4 u^3_\ell \mathcal{O}_{u^4}^{(1)ijkl} + 6 u^2_\ell \mathcal{O}_{u^4}^{(2)ijkl} 
+ 4u_\ell \mathcal{O}_{u^4}^{(3)ijkl}
+\mathcal{O}_{u^4}^{(4)ijkl}\,.
\end{split}
\ee
where $u^i_\ell$, $\delta_\ell$ are the smoothed long-wavelength velocity and density fields (for clarity, we will drop the subscript $\ell$ in the below). To preserve Galilean symmetry, the operators $\mathcal{O}$ should not depend on the smoothed velocity field.

Note that the velocity field scales like $k^{-1}\delta_\k$
at the linear order, \textit{i.e.}\ 
\be
u_i=-\frac{\d_i}{\Delta}\theta \,,
\ee
for Laplacian $\Delta$. Thus, in order to obtain the redshift-space bispectrum counterterms scaling like $k^2\delta^2$,
we need to retain four types of correction: (1) $\mathcal{O}(k^1\delta^2)$ arising from $[u^i(1+\delta_g)]_R$, (2) $\mathcal{O}(k^{0}\delta^2)$ coming from $[u^iu^j(1+\delta_g)]_R$, (3) $\mathcal{O}(k^{-1}\delta^2)$ coming from $[u^iu^ju^k(1+\delta_g)]_R$, and (4) $\mathcal{O}(k^{-2}\delta^2)$ coming from $[u^iu^ju^ku^l(1+\delta_g)]_R$. 

We now discuss each type of composite operator. The renormalized operator $[u^i (1+\delta_g)]_R$ with counterterms $\mathcal{O}(k\delta^2)$ is given by 
\be 
[u^i (1+\delta_g)]_R =u^i (1+\delta)
+\tilde{e}_1 \d_i \delta + \tilde{e}_2 \d_i(s_{kl}s^{kl})+\tilde{e}_3 \d_i \delta^2
+\tilde{e}_4 s_{kl}\d_is^{kl} +\tilde{e}_5 \delta \d_i\delta
\,,
\ee
where $\delta$ is the matter density field and we have introduced the tidal tensor $s_{ij}=[\d_i\d_j \Phi -\delta_{ij}(\Delta \Phi/3)]/(\frac{3}{2}\mathcal{H}^2)$. The renormalized operator $[u^i u^j(1+\delta_g)]_R$ 
with counterterms $\mathcal{O}(k^0\delta^2)$ is given by 
\be 
[u^i u^j(1+\delta_g)]_R = u^i u^j(1+\delta) + 
\tilde{c}_1 \delta_{ij}\delta +
\tilde{c}_2 s_{ij} + c_3 \delta^2 \delta_{ij}
+\tilde{c}_{4}s_{il}s_{lj} \,,\\
\ee
The renormalized operator $[u^i u^j u^k(1+\delta_g)]_R$ with counterterms $\mathcal{O}(k^{-1}\delta^2)$ is given by 
\be 
[u^i u^j u^k (1+\delta_g)]_R = u^iu^ju^k(1+\delta)
+\tilde{b}_1 u_{(i} \delta_{jk)}
+\tilde{b}_2~u_{(i} s_{jk)}
+\tilde{b}_3 \delta u_{(i} \delta_{jk)}\,,
\ee
where $(...)$ denotes symmetrization over indices. Finally, the renormalized operator $[u^iu^ju^ku^l(1+\delta_g)]_R$ 
with counterterms $\mathcal{O}(k^{-2}\delta^2)$ is given by 
\be 
[u^iu^ju^ku^m(1+\delta)]_R =u^iu^ju^ku^m(1+\delta_g)
+\tilde{d}_1 u_{(i} u_j \delta_{kl)} \,.
\ee

In combination, once certain redundant operators are removed, we find the following redshift-space bispectrum counterterm kernel,
\be
 \begin{split}
 & Z_2^{\rm ctr}(\k_1,\k_2) = \Bigg(\bb{C_1} \mu^2 k^2_3 F_2(\k_1,\k_2)  +
\bb{C_2} k_{3z}^2  \left(\mu^2_3-\frac{1}{3}\right) F_2(\k_1,\k_2)   
+ \bb{C_3} k_{3z}^2  \\
&
+\bb{C_{4}} k_{3z}^2 \left(
\frac{ (\k_1\cdot \k_2) \mu_1 \mu_2 }{k_1  k_2} 
-\frac{\mu_1^2+\mu_2^2}{3}  
+\frac{1}{9}\right) 
+ \bb{C_5} f  k_{3z}^2 \mu^2_3  G_2(\k_1,\k_2) \\
& +\bb{C_6} f k_{3z}^3 
\left[\frac{\mu_1}{k_1} 
\left(\mu_2
-\frac{1}{3}\right) + \frac{\mu_2}{k_2} 
\left(\mu_1
-\frac{1}{3}\right)\right]
+\bb{C_7} f  k_z^3  \left[\frac{\mu_1}{k_1}+\frac{\mu_2}{k_2} \right] 
+ \bb{C_8} f^2 k_{3z}^4 \frac{\mu_1}{k_1} \frac{\mu_2}{k_2} \\
& 
 + \bb{C_9} k_{3z}^2 \left(\frac{(\k_1 \cdot \k_2)^2}{k^2_1 k^2_2 }-\frac{1}{3}\right)
 \Bigg)\frac{1}{k_{\rm NL}^2}  \,,
 \end{split}
\ee
involving nine Wilson coefficients, marked in color. The overall bispectrum counterterm is given by 
\be
 \begin{split}
B_{\rm ctr}(\k_1,\k_2,\k_3)=2Z_1(\k_1)Z_1(\k_2)(F_2^{\rm ctr} + Z_2^{\rm ctr})(\k_1,\k_2)P_{11}(k_1)P_{11}(k_2)
+\text{cyclic perms.} \,,
 \end{split}
\ee
where $F_2^{\rm ctr}(\k_1,\k_2)$ is the real space, $\mu$-independent 
kernel. In principle, some of the redshift-space counterterms are the same as coefficients in front of the $\mu$-dependent $k^2P_{11}$ redshift-space counterterms in redshift space, and thus should be constrained by the power spectrum data. Indeed, at leading order we have~\cite{Perko:2016puo}
\be
\delta^{\rm CTR, LO}_z(\k)=  \mathcal{C}_0 k^2\delta+ \mathcal{C}_1 \mu^2 k^2 \delta +\mathcal{C}_2 \mu^4 k^2 \delta\,,
\ee
where 
\be
\mathcal{C}_1 = C_1-\frac{C_2}{3}\,,\quad \mathcal{C}_2 =C_2 + C_5 f\,.
\ee
The relationship between our
power spectrum counterterm coefficients 
($\propto k^2P_L$) and 
$\mathcal{C}_n$ is given by:
\be 
\begin{split}
& c_0 = b_1 \left(\mathcal{C}_0+\frac{\mathcal{C}_1}{3}+\frac{\mathcal{C}_2}{5}\right)
+\frac{\mathcal{C}_0 f}{3}+\frac{\mathcal{C}_1 f}{5}+\frac{\mathcal{C}_2 f}{7}\,,\\
& c_2= \frac{1}{7f}\left (
7 b_1 \mathcal{C}_1+6 b_1 \mathcal{C}_2+7 \mathcal{C}_0 f+6 \mathcal{C}_1 f+5 \mathcal{C}_2 f
\right)
\,,\\
& c_4 = 
\frac{1}{11f^2}\left (
11 b_1 \mathcal{C}_2
+11 \mathcal{C}_1 f
+15 \mathcal{C}_2 f
\right)\,.
\end{split}
\ee


\section{Stochastic terms}\label{app:stoch}

In this section we discuss stochastic contributions to the one-loop bispectrum.

\subsection{Real space}

The stochastic contributions to the galaxy density field in real space are 
given in terms of $\epsilon$ by (some operators are present in~\cite{Perko:2016puo}):
\be
\begin{split}
\delta_\epsilon = d_1\epsilon + d_2 \epsilon \delta 
+\bar d_1\left(\frac{k}{k_{\rm NL}}\right)^2\epsilon 
+ d_2 \left(\frac{k}{k_{\rm NL}}\right)^2 [\epsilon \delta]_\k 
+ d_3[\partial^i\epsilon \partial_i \delta]_\k
+ d_4[\epsilon \Delta \delta]_\k
+ d_5[\Delta \epsilon \delta]_\k\,.
\end{split} 
\ee
Note that $\langle \epsilon \delta \rangle = 0$ by definition. There are two non-trivial possibilities to contract operators in $\delta_\epsilon$ to obtain the tree-level bispectrum contributions (with free coefficients shown in color):
\be
\begin{split} 
& \mathcal{O}\left(
P\bar n^{-1}
\right):\quad \langle b_1\delta_{\k_1} d_2 [\epsilon \delta]_{\k_2} \epsilon_{\k_3} \rangle' = 2\bb{d_2}\frac{\bb{b_1} P(k_1)+\text{cyc.}}{\bar n}\,,\\
& \mathcal{O}\left(
\bar n^{-2}
\right):\quad \langle  \epsilon^3  \rangle' = \frac{\bb{d_1}^3 \bb{A_{\rm shot}}}{\bar n^2}\,,\
\end{split} 
\ee
where primes denote that we drop the Dirac delta function. These match the operators present in the tree-level bispectrum model~\cite{Ivanov:2021kcd}. At the one-loop order we find three distinct contractions:
\be
\begin{split} 
& \mathcal{O}\left( k^2 P\bar n^{-1}\right):\quad \langle b_1\delta_{\k_1} \bar d_2 k_2^2[\epsilon \delta]_{\k_2} \epsilon_{\k_3} \rangle' = \bb{b_1}  \bb{\bar d_2} \frac{ (k_2^2+k_3^2) P(k_1)+\text{cyc.}}{\bar n}\,,\\
& \mathcal{O}\left(k^2 P\bar n^{-1}\right):\quad 
\langle b_{\nabla^2 \delta }k_1^2 \delta_{\k_1}   d_2  [\epsilon \delta]_{\k_2} \epsilon_{\k_3} \rangle' = 
\bb{b_{\nabla^2 \delta }} \bb{d_2} \frac{ k_1^2 P(k_1)+\text{cyc.}}{\bar n}\,,\\
& \mathcal{O}\left(k^2 \bar n^{-2}\right):\quad 
\langle \bar d_1 k^2_1  \epsilon_{\k_1}  \epsilon_{\k_2} \epsilon_{\k_3} \rangle' = 
  \bb{\bar d_1} \bb{A_{\rm shot}} \frac{ k_1^2+\text{cyc.}}{\bar n^2}\,.\\
\end{split} 
\ee
These reproduce the stochastic bispectrum contribution derived in~\cite{Eggemeier:2018qae}.

\subsection{Redshift space}

In redshift space, the density field involves the following terms at tree-level:
\be
\begin{split}
& \delta^{(z)}(\k)= Z_1(\k)\delta^{(1)}(\k)+[Z_2(\delta^{(1)})^2]_\k+d_1\epsilon_\k+  d_2 b_1 [\delta^{(1)} \epsilon]_\k 
-i f k_z d_1 \left[\epsilon \frac{\hat{z}_i\d_i}{\Delta}\theta^{(1)}\right]_\k  \,,
\end{split}
\ee
whilst at next-to-leading order (\textit{i.e.}\ one-loop), we have the additional stochastic contributions,
\be
\begin{split}
& \delta^{(\mu^2k^2)}_{\rm stoch}(\k) = \left(d_1 f\mu^2 k^2 \epsilon + k_z z^i (\epsilon^i + k^i\epsilon)  + k_z^2 z_i z_j \epsilon^{ij} 
\right)(1+\delta_g).
\end{split}
\ee
At leading order we have the LoS-dependent $P\bar n^{-1}$ term derived in~\cite{Ivanov:2021kcd}. To obtain the one-loop expression, we must perform a renormalization of the contact operators involving the stochastic fields, as in Appendix \ref{app:ctrs}. The contact operator renormalization $\mathcal{O}(\epsilon, \epsilon\delta)$ at the leading order in derivatives, gives the following terms, schematically:
\be
\begin{split}
& [u_i(1+\delta_g)]_R\supset u_i^{(1)}\epsilon + (\epsilon_i +k_i \epsilon)(1+\delta_g) 
+\epsilon \d_i \delta + \d_i (\epsilon \delta), \\
& [u_iu_j(1+\delta_g)]_R\supset (\delta_{ij}\epsilon +\epsilon_{ij})(1+\delta) 
+u^{(1)}_i \epsilon_j + u^{(1)}_i k_j \epsilon + s_{ij}\epsilon, \\
& [u_iu_ju_k(1+\delta_g)]_R\supset u^{(1)}_i  \delta_{jk}\epsilon + u^{(1)}_i\epsilon_{jk} .
\end{split}
\ee
These operators generate the following LoS-dependent correction
to the stochastic component of the galaxy density field:
\be
\label{eq:stoch_rsd}
\begin{split}
\delta_{\rm stoch}\supset & k_zu_z^{(1)}\epsilon + k_z[(\epsilon_i z^i +\d_z \epsilon)
(1+\delta_g)] + k_z^2(\epsilon +\epsilon_{zz})(1+\delta_g) \\
& +k_z^2 u^{(1)}_z \epsilon_z + k_z^2 u^{(1)}_z [\d_z \epsilon] 
+ k_z^3(u^{(1)}_z \epsilon + u^{(1)}_z\epsilon_{zz})\\
&+ k_z(\epsilon[\d_z\delta])
+ k_z[\d_z(\epsilon\delta)]+k^2_z[\epsilon s_{zz}]\,.
\end{split}
\ee
Focussing on the $k^2P(k)/\bar n$ corrections to the bispectrum that 
are generated by this density field~\eqref{eq:stoch_rsd}, we find three possibilities:
\begin{enumerate}
    \item Insertion of the redshift-space counterterms $k^2\delta$ and the real-space operator $d_2\delta \epsilon$, \textit{i.e.}\ $\langle k^2\delta [\epsilon \delta]  \epsilon\rangle$:
\be
\begin{split}
& \langle k_1^2 \mu^{2\alpha}_1\delta_{\k_1}   d_2 [\epsilon \delta]_{\k_2} \epsilon_{\k_3} \rangle' =  d_2\frac{ k_1^2 
\mu^{2\alpha}_1 P_L(k_1)+\text{cyc.}}{\bar n}\,,\\
&\langle k_1^2 \mu^{2\alpha}_1\delta_{\k_1} i f k_{2z} d_1 \left[\epsilon \frac{\hat{z}_i\d_i}{\Delta}\theta^{(1)}\right]_{\k_2} \epsilon_{\k_3} \rangle' = - fd_1\frac{k_1^2 \mu_1^{2\alpha+2} P_L(k_1)+\text{cyc.}}{\bar n}\,,
\end{split}
\ee
where $\alpha = 0, 1,2$.

\item Insertion of the redshift-space stochastic counterterm $\sim k^2\epsilon$
and the real-space operator $d_2\delta \epsilon$, \textit{i.e.}\ 
\be
\begin{split}
& \langle Z_1(\k_1)\delta_{\k_1}  d_2 [\epsilon \delta]_{\k_2} k_{3z}^2\epsilon_{\k_3} \rangle' =  d_2\frac{Z_1(\k_1) P_L(k_1) (k_{2z}^2+k_{3z}^2)+\text{cyc.}}{\bar n}\,.
\end{split}
\ee
\item Insertion of the new redshift-dependent stochastic correction $k_z^2 [\delta \epsilon]$, \textit{i.e.}\ 
\be
\begin{split}
& \langle Z_1(\k_1)\delta_{\k_1}  k_{2z}^2[\epsilon \delta]_{\k_2}  \epsilon_{\k_3} \rangle' = \frac{Z_1(\k_1) P_L(k_1) (k_{2z}^2+k_{3z}^2)+\text{cyc.}}{\bar n}\,,\\
& \langle Z_1(\k_1)\delta_{\k_1}  k_{2z}[\d_z\epsilon \delta]_{\k_2}  \epsilon_{\k_3} \rangle' = \frac{Z_1(\k_1) P_L(k_1) k_{2z} k_{3z}+\text{cyc.}}{\bar n}\,,\\
& \langle Z_1(\k_1)\delta_{\k_1}  k^2_{2z}\left[\d_z\epsilon \frac{\d_z\theta}{\Delta}\right]_{\k_2}  \epsilon_{\k_3} \rangle' = -\frac{Z_1(\k_1) P_L(k_1)\mu_1^2 k_{2z} k_{3z}+\text{cyc.}}{\bar n}\,,\\
& \langle Z_1(\k_1)\delta_{\k_1}  k^3_{2z}\left[\epsilon \frac{\d_z\theta}{\Delta}\right]_{\k_2}  \epsilon_{\k_3} \rangle' = \left(
Z_1(\k_1) P_L(k_1)  \frac{k_{1z}}{k_1^2}(k_{2z}^3+k_{3z}^3)+\text{cyc.}
\right)\frac{1}{\bar n}\,,\\
& \langle Z_1(\k_1)\delta_{\k_1}  
k_{2z}
\left[ \epsilon \d_z\delta\right]_{\k_2}  
\epsilon_{\k_3} \rangle' = -\frac{Z_1(\k_1) P_L(k_1)k^2_{1z}+\text{cyc.}}{\bar n}\,,\\
& \langle Z_1(\k_1)\delta_{\k_1}  
k^2_{2z}
\left[ \epsilon \left(\frac{\d_z^2}{\Delta}-\frac{1}{3}\right)\delta\right]_{\k_2}  
\epsilon_{\k_3} \rangle' =  \left(
Z_1(\k_1) P_L(k_1) \left\{\mu_1^2-\frac{1}{3}\right\}(k_{2z}^2+k_{3z}^2)+\text{cyc.}
\right)\frac{1}{\bar n}\,,\\
\end{split}
\ee
\end{enumerate}
Many of these terms happen to be linearly-dependent once 
we apply the bispectrum triangle condition.
All in all, the $\mu$-dependent part of the stochastic
bispectrum at order $k^2 P\bar n^{-1}$ is given by 
\beq
& B^{(k^2 P\bar n^{-1})}_{\rm stoch}(\k_1,\k_2,\k_3)=\Bigg[ \sum_{n=1,2,3,4} 
\bb{S_{n}} k_1^2 \mu^{2(n-1)}_1 
+ \bb{S_5} Z_1(\k_1) (\mu_2^2 k_2^2+\mu_3^2 k_3^2)  
\\\nonumber
& 
+\bb{S_6} Z_1(\k_1)\mu_1^2 (\mu_2^2 k_2^2+\mu_3^2 k_3^2) 
+\bb{S_7}  Z_1(\k_1)\frac{k_{1z}}{k_1^2}(k_{2z}^3+k_{3z}^3) 
 \Bigg]\frac{P_L(k_1)}{k_{\rm NL}^2\bar n}+\text{cyc.}\,,
\eeq
which involves additional nuisance coefficients shown in color.

In addition, there are purely stochastic terms that generate the bispectrum of the order $k^2\bar n^{-2}$. These arise from the following combinations:
\be
\begin{split}
& \langle \epsilon_{\k_1}  \epsilon_{\k_2} k_{3z}^2\epsilon_{\k_3} \rangle' =  \frac{k_{3z}^2+\text{cyc.}}{\bar n^2}\,,
\end{split}
\ee
giving the bispectrum contribution
\be
\begin{split}
& B^{(k^2 \bar n^{-2})}_{\rm stoch}(\k_1,\k_2,\k_3)= \bb{A_{\text{shot},2}} \frac{k_{3z}^2+k_{1z}^2+k_{2z}^2}{\bar n^2}\,.
\end{split} 
\ee
After performing angular integration (in the absence of coordinate-distortion effects), this term takes
the same form as the real-space term $\sim k^2\bar n^{-2}$. Thus, it does not produce a new contribution to the bispectrum monopole, though is important if higher-order multipoles are also considered.

\section{Prior volume effects}
\label{sec:proj}

In this section we study the prior volume effects present in our posteriors when the one-loop bispectrum likelihood is analyzed with small data cuts, such as $\kmax^B=0.12\hMpc$. At face value, the posterior distributions from this analysis are several $\sigma$ away from the true values. However, here we show that as much as half of this shift can be explained by prior volume (marginalization projection) effects. Indeed, such effects are expected to be present when the data volume is not sufficient to tightly constrain model parameters, which is the case for analyses with low $\kmax^B$.

\begin{figure}
\centering
\includegraphics[width=0.99\textwidth]{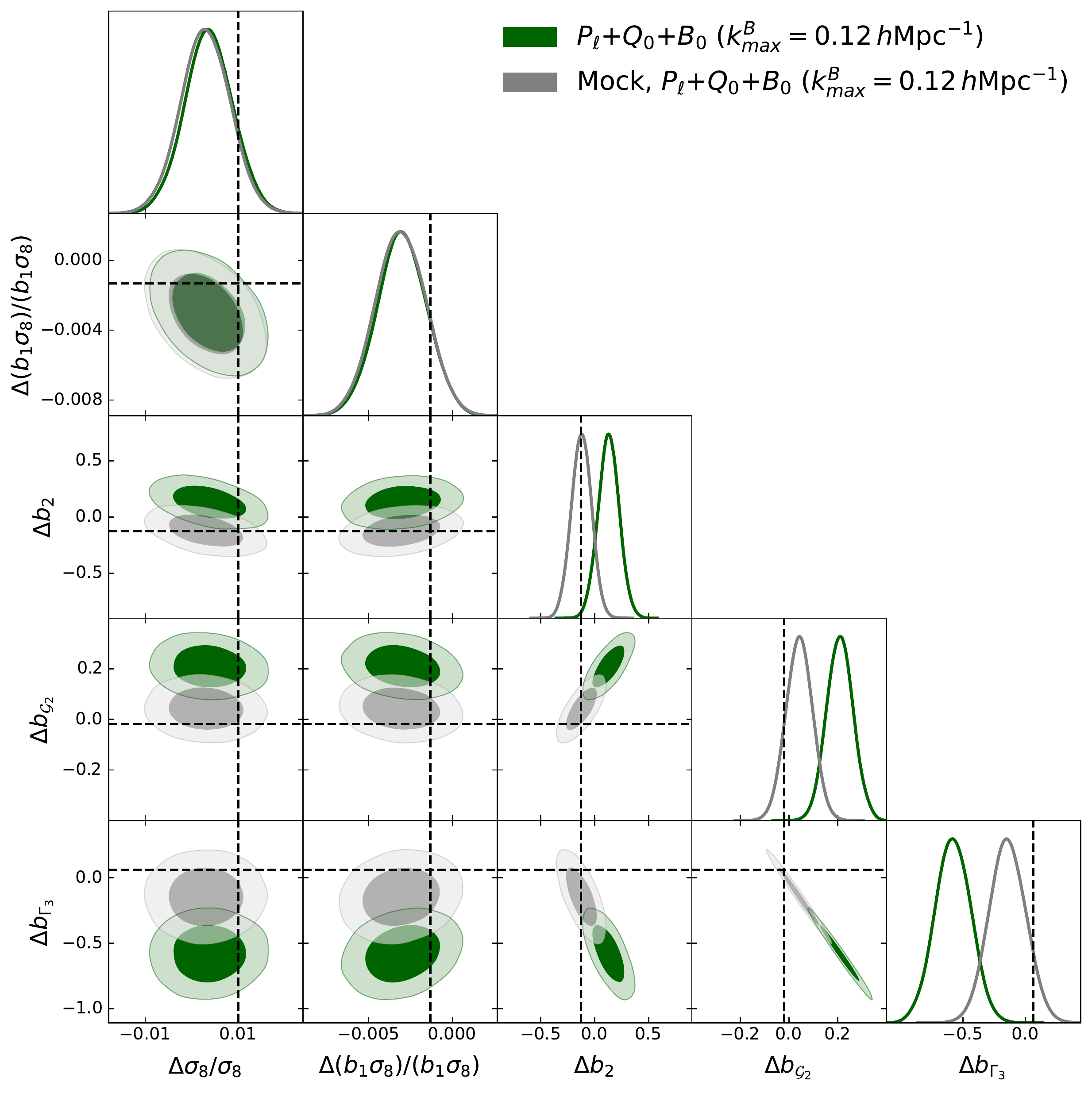}
\caption{
Posterior distributions of the clustering amplitude and low-order nuisance parameters from MCMC analyses of the power spectrum and bispectrum likelihoods from the redshift-space analysis at $\kmax^B=0.12\hMpc$ for the PT Challenge simulation data (in green) and for the mock bispectrum data vector (in gray) computed with our pipeline for the bestfit cosmology at $\kmax^B=0.15\hMpc$. Dashed lines show the input values for the mock data-vector, whose discrepancies with the grey posteriors indicate clear evidence for prior volume effects.} 
\label{fig:prior}
\end{figure}

We performed the following test: rerunning our full analysis on the mock data generated by our fitting pipeline for the best-fit cosmology at $\kmax^B=0.15~\hMpc$. This mock data is simply a theory curve without
any statistical scatter. In the absence of prior volume effects our pipeline 
must exactly recover the input parameters. However, when we fit this mock bispectrum data at $\kmax^B=0.12~\hMpc$, we find that the mean values recovered from our pipeline are shifted relative to the input values at the $(1-1.5)\sigma$ level, as shown in Fig.\,\ref{fig:prior}. This is evidence of prior volume effects. Furthermore, the shifts are in the directions of the apparent biases observed in the actual data (\S\ref{sec: results-rsd}). Thus, if we subtract these shifts from the actual posteriors at $\kmax^B=0.12$, the mean posterior values would match the true input parameter values at least within the $99\%$ CL. Finally, we note that the parameters $b_2,b_{\mathcal{G}_2}$ and $b_{\Gamma_3}$
are highly correlated; this means that a shift in one would induce a shift in both.

As an additional check, we repeat our mock analysis for $\kmax^B=0.15\hMpc$. 
Overall, we find much improved agreement between the mock and actual analyses.  
The posteriors for $\sigma_8$ and $b_{\Gamma_3}$
are still shifted with respect to the ground truth 
by $\lesssim 1\sigma$ (which is smaller than $1.5\sigma$ shifts in the $\kmax^B=0.12\hMpc$ case), but all other parameters
are recovered without noticeable bias.

\begin{figure}
\centering
\includegraphics[width=0.99\textwidth]{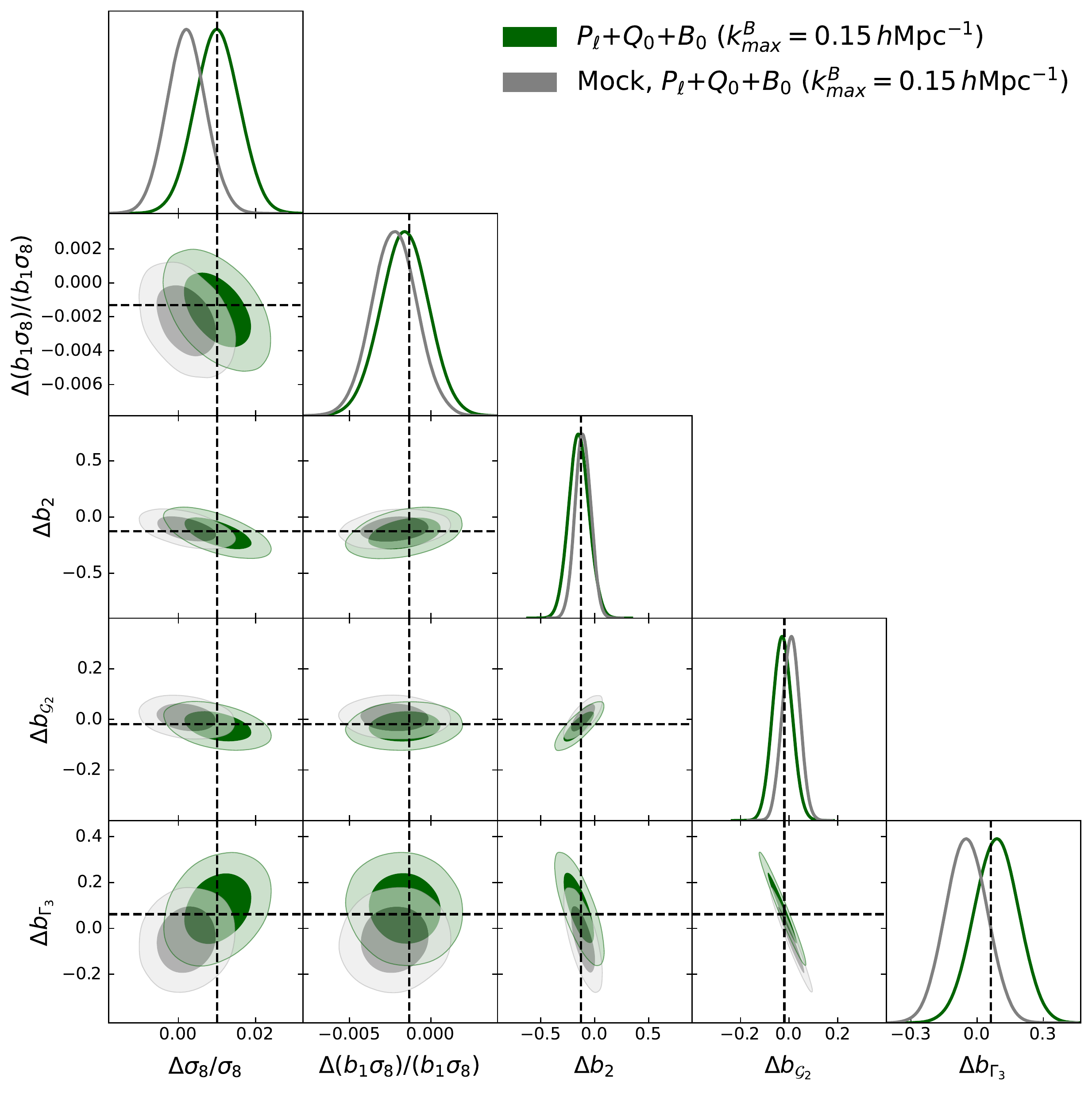}
\caption{As Fig.\, \ref{fig:prior}, but for $\kmax^B=0.15\hMpc$.
}
\label{fig:prior15}
\end{figure}

\bibliography{short}
\bibliographystyle{JHEP}

\end{document}